\documentclass[prd,nofootinbib,preprint]{revtex4}
\usepackage[utf8]{inputenc}
\usepackage{amsmath}
\usepackage{amssymb}
\usepackage{textcomp}
\usepackage{array,multirow}
\usepackage{slashed}
\usepackage{tabularx}
\usepackage{hyperref}
\usepackage{graphicx}
\usepackage{caption}
\usepackage{caption}
\usepackage{ragged2e}
\usepackage{xcolor}
\usepackage{textcomp}
\usepackage{cancel}
\usepackage[labelformat=parens]{subcaption}
\usepackage{varwidth}

\begin{document}
\title{Baryogenesis via flavoured leptogenesis in a minimal type-II seesaw 
model}
\author{Sreerupa Chongdar}
\email{518PH1002@nitrkl.ac.in}
\affiliation{Department of Physics and Astronomy, 
National Institute of Technology Rourkela, Sundargarh, Odisha, India, $769008$}
\author{Sasmita Mishra}
\email{mishras@nitrkl.ac.in}
\affiliation{Department of Physics and Astronomy, 
National Institute of Technology Rourkela, Sundargarh, Odisha, India, $769008$}

\begin{abstract}
 We study baryogenesis via leptogenesis in an extension of the Standard Model by adding one right-handed neutrino and
one triplet scalar. These heavy particles contribute to the
generation of tiny neutrino mass through seesaw mechanism. The contribution of the heavy particles to the neutrino masses 
is inversely proportional
to their corresponding masses. Considering 
leptogenesis is achieved by the decay of the right-handed neutrino,
the new source of CP asymmetry comes solely from the decay of the 
right-handed neutrino with one-loop vertex diagram involving 
the triplet scalar. The predictiveness of the model is  enhanced
by introducing Fritzsch-type textures for
the neutrino mass matrix and charged lepton
mass matrix. We execute the parameter space study 
following the latest neutrino oscillation data.
We study baryogenesis via leptogenesis in the two-flavoured regime, using the 
zero textures, and show that there is an enhancement in baryon asymmetry as compared to the unflavoured regime.
For two-flavour leptogenesis we consider the suitable temperature regime $T\subset\left[10^{10},10^{11}\right]$ GeV. We also study the common correlation of CP violation between low and high-energy
regimes using the geometrical description of CP violation in terms of
unitarity triangle.
\end{abstract}

\maketitle

\section{Introduction}
\label{sec:Intro}
The two enigmatic problems of non-zero neutrino mass and baryon asymmetry of the Universe
find a common solution in the Standard Model
(SM) augmented with a certain choice of additional heavy fields. Through the seesaw mechanism
one can provide a possible theoretical explanation of nonzero neutrino mass,
confirmed by the experimental observation of
neutrino flavour oscillation \cite{Giunti:2007ry}. The observed baryon
asymmetry of the Universe (BAU) could also be explained through leptogenesis
\cite{Fukugita:1986hr} via out-of-equilibrium decay of the same heavy fields taking part in the seesaw
mechanism. The former is one low-energy observation
(after electroweak symmetry breaking) whereas the 
latter, a high-energy phenomenon (before electroweak symmetry breaking).
Thus seesaw mechanism provides a nontrivial link between the generation of 
light neutrino mass and baryogenesis through leptogenesis. In this work,
we study the generation of BAU through flavoured leptogenesis in the minimally extended SM by 
one fermion singlet  and one scalar triplet. The phenomena of CP violation
is inevitable both in neutrino oscillation and leptogenesis. We also establish a link between
low- and high-energy CP violations.

One of the famous frameworks of neutrino mass generation via 
seesaw mechanism and leptogenesis via out-of-equilibrium decay of 
a heavy beyond SM field is the one where the SM is extended with heavy right-handed
neutrinos. One needs more than just one right handed neutrino to 
account for light neutrino mass generation, compatible with experimental 
data and leptogenesis. 
The addition of right-handed neutrinos is also consistent
with the theories inspired by Grand Unification such as 
Left-right symmetry \cite{Mohapatra:1974gc,Senjanovic:1975rk,Senjanovic:1978ev},
Pati-Salam \cite{Pati:1974yy} and SO$(10)$ \cite{Georgi:1974my,Fritzsch:1974nn}. 
However, in such theories, heavy fields such as scalar triplets and 
fermion triplets arise naturally and can establish the connection between 
light neutrino mass generation via seesaw mechanism and leptogenesis.

Keeping minimal extension of the SM in mind, 
we study a minimal type-II seesaw model, where the SM is extended with
one $SU(2)_L$ triplet scalar, $\Delta$ and one right-handed singlet fermion, $N$.
So in our study, there are two mass scales involved: the mass of
the right-handed singlet, $M$ and that of scalar triplet, $M_\Delta$. 
While both the fields contribute to light neutrino mass via seesaw
mechanism, considering hierarchical mass limits, the out-of-equilibrium decay of one field is responsible
for creating lepton asymmetry and hence BAU via leptogenesis. 
The requirement of CP violation being essential for baryogenesis via leptogenesis \cite{Sakharov:1967dj}, it is possible to produce CP violation in two ways in such a model. The vertex diagram
involving one right-handed neutrino and one triplet scalar can be of 
two types, depending on the relative hierarchy of their masses:
\begin{enumerate}
 \item  If we have the triplet scalar as the lightest seesaw state 
 $\left(M_\Delta \ll M\right)$, then the decay of the triplet scalar dominates
 the CP asymmetry in the presence of a virtual right-handed neutrino
 present in the vertex diagram.
 \item If the right-handed neutrino is lighter than the triplet 
 scalar $\left(M\ll M_\Delta\right)$, the CP asymmetry is produced predominantly
 from the decay of the right-handed neutrino in the presence of a 
 virtual triplet scalar in the vertex diagram. 
 \end{enumerate}
In the SM, considering there are not so light scalars we choose 
to work in the hierarchy $M\ll M_\Delta$. In a model
with $n$ number of right-handed neutrinos and one triplet scalar,
the neutrino mass can be generated from two sources,
\begin{equation}
 m_{\nu}= m^{(I)}_{\nu}+ m^{(II)}_{\nu},
\end{equation}
where $m^{(I)}_{\nu}$ is the right-handed neutrino contribution 
coming from the type-I seesaw mechanism, and $m^{(II)}_{\nu}$ is 
the triplet scalar contribution coming from the type-II seesaw
mechanism. If we consider
the lightest right-handed neutrino $N_{1}$ to be the lightest seesaw
state, then the CP asymmetry can be obtained from the decay
of $N_{1}$ in the presence of the other right-handed neutrinos
$\left(N_{2},...N_{n}\right)$ or the triplet scalar $\Delta$ in the
vertex diagram. There is no one-to-one correspondence
by each seesaw state, between the amount of contribution to neutrino mass 
and CP asymmetry (as the neutrino mass matrix is a $3\times 3$ complex matrix, not a number). But the contribution of the seesaw states to the 
production of CP asymmetry is found to be proportional to 
their respective contribution to the neutrino mass generation \cite{Hambye:2003ka}.
In that case, in the limit, $m^{(I)}_{\nu}\gg m^{(II)}_{\nu}$, the contribution of the triplet 
scalar in the production of CP asymmetry can be safely ignored. 
However, in a minimal type-II seesaw model, 
the CP asymmetry can be obtained only from the one-loop 
vertex diagram involving the triplet scalar as there is only 
one right-handed neutrino. So we will consider the contribution 
of triplet scalar in the CP asymmetry production.

In the study of baryogenesis through leptogenesis flavour effects are known to induce some 
novel feature as compared to unflavoured case especially due to the nature of 
wash-out effects along different directions in flavour space. 
At higher temperature, $T\gtrsim 10^{12}$ GeV, the charged lepton
flavours $\left(e, \mu, \tau\right)$ are out of thermal equilibrium and thus
indistinguishable. In this case, the leptogenesis can be successfully 
expressed in an unflavoured regime. However, at a temperature 
$T\lesssim 10^{12}$ GeV, the processes induced by $\tau$-Yukawa 
are in thermal equilibrium. It breaks the coherence between 
$\tau$-lepton and the other two leptons $\left(e, \mu\right)$. The lepton 
asymmetries are expressed through $Y_{\Delta_{a}}$ and 
$Y_{\Delta_{\tau}}$ in this temperature range, where $a=e+\mu$ 
is the superposition of the flavours $e$ and $\mu$. Further 
below $T\sim 10^{9}$ GeV, the interactions induced by the 
$\mu$-Yukawa are in thermal equilibrium, completely breaking 
the flavour coherence. So, leptogenesis needs to be 
studied in terms of fully flavoured lepton asymmetries, 
$Y_{\Delta_{e}}$, $Y_{\Delta_{\mu}}$, and $Y_{\Delta_{\tau}}$ 
at temperature $T\lesssim 10^{9}$ GeV. 

In seesaw models, generally, there are many free parameters as long as
the coupling matrices are concerned. There are not
enough experimental constraints to fix the parameters. The number of
free parameters can be reduced by imposing texture zeros in the  coupling matrices.
Often they are motivated by imposing new symmetries on the particle content
of the model. In this way, the model becomes predictive by fitting the parameters
to low-energy neutrino data. Taking Fritzsch-type textures  \cite{Fritzsch:1979zq,Xing:2002sb,Fukugita:2003tn,Xing:2004hv, Fritzsch:2011qv} into account we
show that the BAU can be enhanced to comply with observational value, considering flavoured effects as
compared to unflavoured case. This feature is helpful in bringing down the scale of leptogenesis. We study the two-flavoured leptogenesis
in the temperature range $T\sim\left[10^{10}, 10^{11}\right]$ GeV. 
We find that the lepton 
asymmetries obtained through leptogenesis can lead to baryon 
asymmetry of the order $\sim 10^{-10}$, and the proper flavour 
consideration enhances the production of baryon asymmetry. It is believed that the CP violation in low-energy (e.g. neutrino oscillation) and high-energy regimes (e.g. leptogenesis)
are in general not related. Using a geometrical interpretation of CP violation
at low-energy, we also show a common link between low and high
energy CP violation in this setup. 

The paper is organized as follows. In section (\ref{sec-Nmm}), a minimal type-II neutrino mass model is 
introduced. In section (\ref{sec-AnR}), a detailed parameter space study 
of the neutrino mass matrix elements is carried out, which covers 
the diagonalization of matrices and the interesting correlation 
among different mixing angles arising from the model. The allowed 
parameter space indicated in this section is further used to obtain
CP asymmetry parameters. In section (\ref{sec-BnL}), the processes like 
baryogenesis and leptogenesis are discussed from a cosmological
point of view. Section (\ref{sec:result-bau}) contains the results and
discussions. It gives a comparative study among these regimes of 
leptogenesis, generically judged based on different right-handed 
neutrino masses $M$. The common origin of CP violations in low- and high-energy sectors is analyzed in section (\ref{sec:cp-low-high}).
 Finally, in section (\ref{sec-Conclusion}), we give our conclusions of the study. 
\section{Neutrino mass model}
\label{sec-Nmm}
 With addition of a triplet scalar, $\Delta$ and $n$-number of right-handed
 Majorana neutrinos, $N_{i}$ $\left(i=1,....,n\right)$, the extended SM Lagrangian can 
 be written as
 \begin{equation}
  -\mathcal{L}\supset {Y_{l}}_{\alpha \beta} \overline{L}_{\alpha}{l_R}_{\beta}\phi+ {Y_\nu}_{\alpha i}\overline{L}_{\alpha}N_{i}\phi+ 
  \frac{1}{2}M_{i} \bar{N}_{i} N_{i} +Y_{\Delta_{\alpha\beta}} \bar{L}^{C}_{\alpha} i\tau_{2}\Delta L_{\beta}- \mu\phi^{T}i\tau_{2}\Delta\phi +M^{2}_{\Delta}{\rm Tr}\Delta^{\dagger}\Delta +{ \rm h.c.},
 \end{equation}
where $L_{\alpha}=\left(\nu_{\alpha},l_{\alpha}\right)^{T}$ and ${l_R}_{\beta}$, $\left(\alpha, \beta =e,\mu,\tau\right)$
are the left- and right-handed SM leptons respectively. The SM Higgs doublet is represented as $\phi=\left(\phi^{0},\phi^{-}\right)^{T}$. The triplet scalar, $\Delta$ can be represented in $SU(2)_L$ adjoint representation as,
\begin{equation*}
 \Delta=
 \begin{pmatrix}
  \frac{\delta^{+}}{\sqrt{2}} & \delta^{++}\\
  \delta^{0} & -\frac{\delta^{+}}{\sqrt{2}}
 \end{pmatrix}.
\end{equation*}
The Dirac-type Yukawa coupling matrix of the right-handed neutrinos and charged leptons with the SM leptons and Higgs scalar are represented as $Y_\nu$ and $Y_l$ respectively.
The coupling matrix, $Y_\Delta$ is a $3\times3$ Majorana-type Yukawa 
coupling matrix of the triplet scalar with the SM lepton doublets. After electroweak symmetry breaking, due to 
the vacuum expectation value (vev) developed by the neutral component of 
the doublet Higgs, $v = \langle \phi_0 \rangle$,
the neutral component of $\Delta$ also 
acquires a vev, $v_{\Delta} =\langle \delta_0 \rangle \simeq\frac{\mu^{*}
v^{2}}{M^{2}_{\Delta}}$. The triplet vev, $v_\Delta$ is seesaw suppressed 
for heavy triplet scalar. Once the heavy degrees of freedom are integrated out, as a consequence, there are two sources of light neutrino masses,
\begin{equation}
 m_{\nu}=m^{(I)}_{\nu}+m^{(II)}_{\nu}=-Y_\nu^{*}\frac{1}{M}Y_\nu^{\dag}v^{2}+2Y_\Delta v_{\Delta}.
  \label{eq:neumass}
\end{equation}
The first and second terms are mass terms due to type-I and II seesaw induced masses
respectively, with $v= 174$ GeV. 
 The low-energy neutrino oscillation experiments provide data for six parameters:  
three mixing angles, two mass-squared differences, and one CP violation phase. 
In our model, the Dirac-type Yukawa coupling for right-handed neutrinos has
$6$ ($3$ moduli and $3$ phases for one right-handed neutrino) and the Majorana-type Yukawa
which is complex symmetric, has $12$ ($6$ moduli and $6$ phases) independent parameters. 
In order to make the model predictive we assume the Fritzsch-type textures \cite{Fritzsch:1979zq,Xing:2002sb,Fukugita:2003tn,Xing:2004hv, Fritzsch:2011qv}
of charged lepton mass matrix, $m_{l}$ and triplet scalar induced neutrino mass matrix, $m_{\nu}^{(II)}$. Following the same texture for both,
\begin{equation}
 m_{l}=v
  \begin{pmatrix}
   0 & C_{l}e^{i\alpha_{l}} & 0\\
   C_{l}e^{i\alpha_{l}} & 0 & B_{l}e^{i\beta_{l}}\\
   0 & B_{l}e^{i\beta_{l}} & A_{l}e^{i\gamma_{l}}
  \end{pmatrix},
  \label{eqn-oml}
 \end{equation}
 and
 \begin{equation}
m^{(II)}_{\nu}=v_{\Delta}
  \begin{pmatrix}
   0 & C_{\nu}e^{i\alpha_{\nu}} & 0\\
   C_{\nu}e^{i\alpha_{\nu}} & 0 & B_{\nu}e^{i\beta_{\nu}}\\
   0 & B_{\nu}e^{i\beta_{\nu}} & A_{\nu}e^{i\gamma_{\nu}}
  \end{pmatrix}.
  \label{eq:m-nu-ii}
 \end{equation}
 Such textures arise in left-right symmetric models, where right-handed neutrinos and
 triplet scalars arise naturally and were studied in Ref. \cite{Borgohain:2019pya}. Also,
 in the construction of models based on Froggatt-Nielsen mechanism \cite{Froggatt:1978nt}, texture zeros arise in the neutrino Yukawa coupling due to the assignment of different charges under an additional symmetry to particles of different generations. The consequences of texture zeros in $(1,1), (1,2)$ and $(1,3)$ positions of Yukawa matrix have interesting consequences in flavoured leptogenesis as shown in Ref.\cite{Abada:2006ea}. For example in the case of texture zero in $(1,1)$ position, $e$-CP asymmetry is weakly washed out while the $\mu$- and $\tau$-CP asymmetries
 are strongly washed out. In the quark sector, the simultaneous presence of 
 zeros in the $(1,1)$ elements of symmetric up and down quark mass matrices leads to the prediction of Cabibbo angle $\theta_C\simeq \sqrt{m_d/m_s}$ \cite{Gatto:1968ss}.

In our case, we have one right-handed neutrino. So the corresponding Yukawa coupling matrix is a
column matrix, which can be set as \cite{Gu:2006wj},
 \begin{equation}
  Y_\nu = i y_0 \left(0, r, 1\right)^T.
  \label{eq:yukawa}
 \end{equation}
The purpose of introducing imaginary unit $i$
is to cancel the minus sign of the type-I term 
for convenience. The appearance of zero in the $(1,1)$ position of the
above Yukawa matrix ensures two-zero texture in the total light neutrino
mass matrix as can be seen in the subsequent equation. Now using the equations (\ref{eq:m-nu-ii}) and (\ref{eq:yukawa}) 
in Eq.(\ref{eq:neumass}), the neutrino mass matrix turns out to be
\begin{equation}
  m_{\nu}=m_{0}
  \begin{pmatrix}
   0 & \hat{C_\nu}e^{i\alpha_{\nu}} & 0\\
   \hat{C_\nu}e^{i\alpha_{\nu}} & r^{2} & r+\hat{B_\nu}e^{i\beta_{\nu}}\\
   0 & r+\hat{B_\nu}e^{i\beta_{\nu}} & 1+\hat{A_\nu}e^{i\gamma_{\nu}}
  \end{pmatrix}.
  \label{eqn-omnu}
 \end{equation}
 Here $m_0 \equiv v^2 y_0^2/M$ and $\hat{A_{\nu}}\equiv v_{\Delta}A_{\nu} / m_{0}$ 
 and similarly for $\hat{B_\nu}$ and $\hat{C_\nu}$.
\subsection{Parameter space determination by confronting with neutrino data}
\label{sec-AnR}
To study the neutrino mass generation and to produce optimum CP asymmetry from
the neutrino mass model, we need to study the parameter space offered by the model.

There are three important steps to be followed for diagonalizing the matrices 
and make a connection with experimental observations \cite{Xing:2003zd,Nishiura:1999yt}.
\begin{enumerate}
 \item In the first step the charged and neutral lepton mass matrices are
 decomposed in terms of  diagonal phase matrix, $P_{l,\nu}$ and real 
 symmetric matrix, $\bar{m}_{l,\nu}$ so that
 \begin{equation}
  m_{l,\nu} = P^T_{l,\nu} \bar{m}_{l,\nu} P_{l,\nu},
  \label{eqn-mPmP}
 \end{equation}
where
\begin{equation}
 P_{l,\nu} = 
 \begin{pmatrix}
    e^{i\theta_{l,\nu}} & 0 & 0\\
    0 & e^{i\phi_{l,\nu}} & 0\\
    0 & 0 & e^{i\psi_{l,\nu}}
   \end{pmatrix}.
\end{equation}
\item In the second step the real symmetric matrix, $\bar{m}_{l,\nu}$ is diagonalized 
following unitary transformation:
\begin{equation}
 U^T_l \bar{m}_{l} U_l = {\rm Diag}\left(m_e,m_\mu,m_\tau\right),~ U^T_\nu \bar{m}_{\nu} U_\nu =
 {\rm Diag}\left(m_1,m_2,m_3\right),
\end{equation}
where $m_e, m_\mu$ and $m_\tau$ are the masses of $e, \mu$ and $\tau$
leptons respectively. The mass eigenvalues of the light neutrino mass
matrix are represented as $m_1, m_2$ and $m_3$.
\item The lepton flavour mixing matrix, $V$ then arises from the mismatch between 
the diagonalization of the charged and neutral mass matrices: $V= U^T_l \left(P^*_l  P_\nu\right) 
U_\nu^*$. The elements of the matrix can be written as
\begin{equation}
 V_{pq}={U_l}_{1p}{U_\nu}^{ *}_{1q}e^{i\theta}+{U_l}_{2p}{U_\nu}^{*}_{2q}e^{i\phi}
 +{U_l}_{3p}{U_\nu}^{ *}_{3q}e^{i\psi},
 \label{eqn-vpq}
\end{equation}
where $p\equiv\left(e,\mu,\tau\right)$, $q\equiv\left(1,2,3\right)$. The phases are defined as, 
$\theta=\left(\theta_{\nu}-\theta_{l}\right)$, $\phi=\left(\phi_{\nu}-\phi_{l}\right)$, $\psi=\left(\psi_{\nu}-\psi_{l}\right)$. The 
elements of $U_\nu$ and $U_l$ are given in the appendices (\ref{app-1}) and (\ref{app-2}).
\item The elements of the mixing matrix $V$ depends on only two combinations
of three phases, $\left(\theta, \phi, \psi\right)$ as the overall phase of $V$ has 
nothing to do with experimental observable. The elements of the matrices 
$U_{l,\nu}$ also depend on the mass ratios of the charged and neutral 
leptons, as given below,
\begin{equation}
 x_{l}=\frac{m_{e}}{m_{\mu}}, \quad 
 y_{l}=\frac{m_{\mu}}{m_{\tau}},
\end{equation}
\begin{equation}
 x_{\nu}=\frac{m_{1}}{m_{2}}, \quad 
 y_{\nu}=\frac{m_{2}}{m_{3}}.
 \label{eqn-xyn}
\end{equation}
The charged lepton ratios $x_{l}$, $y_{l}$ are now determined with better accuracy, 
as 
\begin{equation}
 x_{l}\simeq 0.00484, \quad y_{l}\simeq 0.0594.
 \label{eqn-xyl}
\end{equation}
So, there are only four free parameters (two phases and $x_\nu$ and $y_\nu$) that
can be constrained from neutrino oscillation data.
\end{enumerate}

\begin{figure}[t]
\includegraphics[width=14.6cm]{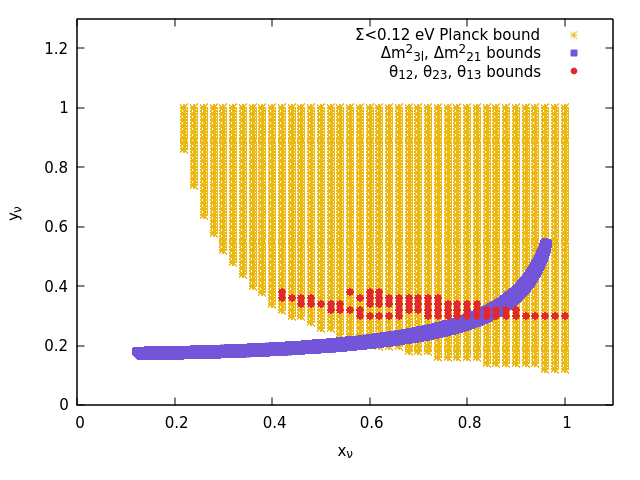} 
\caption{The figure is a representation of the determination of parameter space of $[x_{\nu},y_{\nu}]$. 
 The blue region is bounded by the constraints
coming from two mass-squared differences, 
$\frac{\Delta m^{2}_{21}}{10^{-5}}( eV^{2})\subset [6.82,8.04]$, 
and $\frac{\Delta m^{2}_{31}}{10^{-3}} (eV^{2})\subset[2.430,2.593]$. 
The yellow region is bounded by the Planck constraint on the total neutrino mass 
$\Sigma=m_{1}+m_{2}+m_{3}<0.12$ eV.
The red region is bounded by the constraints coming from the three mixing 
angles $\theta_{12}\subset[31.27^{\circ},35.87^{\circ}]$,
$\theta_{23}\subset[39.7^{\circ},50.9^{\circ}]$, and
$\theta_{13}\subset[8.25^{\circ},8.98^{\circ}]$. The 
intersection of these three regions is the small allowed 
parameter space for $x_{\nu}\subset[0.80-0.90]$ and $y_{\nu}\subset[0.30-0.35]$.}
\label{fig-xytst3}
\end{figure}

The three mixing angles $\theta_{ij}$ of the neutrino oscillation parameters can 
be expressed in terms of the lepton flavour mixing matrix $V$, as follows:
\begin{equation}
  \sin^{2}2\theta_{12}=4\left|V_{e1}\right|^{2}\left|V_{e2}\right|^{2},
 \end{equation}
\begin{equation}
 \sin^{2}2\theta_{23}=4\left|V_{\mu 3}\right|^{2}\left(1-\left|V_{\mu 3}\right|^{2}\right),
\end{equation}
\begin{equation}
 \sin^{2}2\theta_{13}=4\left|V_{e3}\right|^{2}\left(1-\left|V_{e3}\right|^{2}\right).
\end{equation}
The parameter space of $\left[x_{\nu},y_{\nu}\right]$ can be restricted by the bounds on mixing angles. 
The experimental constraints are given by \cite{Esteban:2020cvm},
\begin{equation}
 31.27^{\circ}<\theta_{12}<35.87^{\circ},~
 39.7^{\circ}<\theta_{23}<50.9^{\circ},~
 \label{eqn-angles}
 \end{equation}
\begin{equation}
 8.25^{\circ}<\theta_{13}<8.98^{\circ},~
 144^{\circ} < \delta < 350^{\circ}.
\end{equation}
Even though the Dirac CP phase $\delta$ is bounded as
\begin{equation}
144^{\circ} < \delta < 350^{\circ},
\end{equation}
we have used the full range of $\delta\sim[0:360^{\circ}]$ for the parameter space study.
The latest bounds on the two mass-squared differences are given by 
\begin{equation*}
 6.82\times10^{-5}~ {\rm eV}^{2}<\Delta m^{2}_{21}<8.04\times10^{-5}~ {\rm eV}^{2},
\end{equation*}
\begin{equation}
 2.430\times10^{-3} ~{\rm eV}^{2}<\Delta m^{2}_{31}<2.593\times10^{-3} ~{\rm eV}^{2}.
 \label{eqn-mass-sqr}
\end{equation}
Once the values of $x_\nu$ and $y_\nu$ are constrained the absolute
values of three neutrino masses can be found by using
the mass-squared differences, 
\begin{equation}
 \Delta m^{2}_{21}=m^{2}_{2}-m^{2}_{1} = m_2^2\left|1-x_\nu^2\right|,\quad
  \Delta m^{2}_{31}=m^{2}_{3}-m^{2}_{1} = m_3^2 \left| 1-y_\nu^2\right|.
 \label{eqn-dm213l}
\end{equation}
In order to diagonalize the charged lepton and neutrino mass matrices given in 
equations (\ref{eqn-oml}) and (\ref{eqn-omnu})
and determine the parameter space using experimental data we follow the steps 
laid above. In order to write the lepton mass matrices $m_l$ and $m_\nu$ in factorized form as given in Eq.(\ref{eqn-mPmP}), we make two assumptions here: 
$r= \sqrt{m_2/m_0}$ and ${\rm arg} \left(1+\hat{A_\nu} e^{i\gamma_\nu}\right) = 2 {\arg}
\left(r+ \hat{B_\nu}\right) e^{i\beta_\nu}$ \cite{Xing:2003zd} and then obtain 
\begin{equation}
  \hat{A_{\nu}}=\left[\frac{(m_{3}-m_{1})^{2}}{m^{2}_{0}}-\sin^{2}
  \gamma_{\nu}\right]^{\frac{1}{2}}-\cos\gamma_{\nu},
  \label{eqn-Ahat}
 \end{equation}
 \begin{equation}
  \hat{B_{\nu}}=\left[\frac{m_{1}m_{3}(m_{3}-m_{1}-m_{2})}
  {m^{2}_{0}(m_{3}-m_{1})}-r^{2}\sin^{2}\beta_{\nu}\right]^{\frac{1}{2}}-r\cos\beta_{\nu},
  \label{eqn-Bhat}
 \end{equation}
 \begin{equation}
  \hat{C_{\nu}}=\left[\frac{m_{1}m_{2}m_{3}}{m^{2}_{0}(m_{3}-m_{1})}\right]^{\frac{1}{2}}.
  \label{eqn-Chat}
 \end{equation}
 Similarly, the elements of the charged lepton mass matrix, 
 shown in Eq.(\ref{eqn-oml}), can be expressed in terms of
 three charged lepton masses $m_{e}$, $m_{\mu}$ and $m_{\tau}$.
\begin{equation}
 A_{l}=\left(m_{\tau}-m_{\mu}+m_{e}\right),
\end{equation}
\begin{equation}
 B_{l}=\left[\frac{(m_{\mu}-m_{e})(m_{\tau}-m_{\mu})(m_{e}-m_{\tau})}
 {(m_{\tau}-m_{\mu}+m_{e})}\right]^{\frac{1}{2}},
\end{equation}
\begin{equation}
 C_{l}=\left[\frac{m_{e}m_{\mu}m_{\tau}}{(m_{\tau}-m_{\mu}+m_{e})}\right]^{\frac{1}{2}}.
\end{equation}
In order to determine the parameter space of $x_\nu$ and $y_\nu$
we assume the general range of the lightest neutrino mass in normal hierarchy (NH), 
as $m_{1}\sim\left[0.001:0.05\right]$ eV. Hence, the other neutrino masses can be 
calculated directly from Eq.(\ref{eqn-dm213l}). The plot in Fig.(\ref{fig-xytst3}) shows how different bounds restrict
the parameter space of $x_\nu$ and $y_\nu$.  In Fig.(\ref{fig-xytst3}), the scattered points
 represent the allowed values of $x_\nu$ and $y_\nu$ that satisfy 
total neutrino mass $\Sigma=m_{1}+m_{2}+m_{3}<0.12$ eV \cite{Planck:2018vyg}, mass-squared differences and mixing angles as given in equations (\ref{eqn-angles}) - (\ref{eqn-mass-sqr}).
One can see a commonly allowed parameter space as the allowed ranges of $x_{\nu}$ and $y_{\nu}$ are given by,
\begin{equation}
 x_{\nu}\sim\left[0.80-0.90\right],\quad y_{\nu}\sim\left[0.30-0.35\right].
 \label{eqn-xyrnge}
\end{equation}
We shall use these ranges to estimate the CP asymmetry parameter for 
the study of leptogenesis.
The parameter space for the neutrino mass matrix elements $\hat{A_{\nu}}$, 
$\hat{B_{\nu}}$ and $\hat{C_{\nu}}$ can be further constrained using the latest 
neutrino oscillation parameter data using equations (\ref{eqn-Ahat} - \ref{eqn-Chat}), 
as the model contributes to the formation 
of baryon asymmetry through these matrix elements. 
\section{Baryogenesis and Leptogenesis}
\label{sec-BnL}
The baryon asymmetry of the Universe, $\eta_{B}= \left(n_B-n_{\overline{B}}\right)/n_\gamma$ is 
constrained from the Big Bang Nucleosynthesis (BBN) and Cosmic Microwave Background Radiation (CMBR) 
 data \cite{Amsler:2008zzb}, and is  given by, 
\begin{equation}
 4.7\times 10^{-10}\leq\eta_{B}\leq 6.5\times 10^{-10}, 
 \label{eqn-bbn}
\end{equation}
where $n_{B}$, $n_{\overline{B}}$,  and $n_{\gamma}=\frac{2\zeta(3)}{\pi^{2}}T^{3}$ denote 
the number densities of baryons, antibaryons, and gamma photons respectively.
Theoretically, baryogenesis through leptogenesis \cite{Strumia:2006qk} provides 
a framework for generating the required BAU following three Sakharov's conditions \cite{Sakharov:1967dj}. In the seesaw model under consideration, they are satisfied as follows: 
(i)  The lepton number violation comes into the scenario from the decay of right-handed neutrino considering neutrinos 
as Majorana particles. (ii) CP violation is ensured from the interference of the tree-level and one-loop diagrams of the decay of the
right-handed neutrino. (iii) The out-of-equilibrium conditons is achieved by determining the particle asymmetries, while considering the decay and inverse decay processes in
an expanding early Universe through a set of Boltzmann equations. The lepton asymmetry is converted to the baryon asymmetry through $B+L$ violating Sphaleron processes \cite{Khlebnikov:1988sr}.

If the temperature
is as high as $T\sim10^{12}$ GeV or more, the individual flavours of the leptons do not appear 
to be of much importance, and in this case, solving a set of flavour-independent or 
unflavoured Boltzmann equations proves to be sufficient to study the leptogenesis. 
On the other hand, in the temperature range $T\subset\left[10^{10},10^{12}\right]$ GeV, $\tau$-Yukawa 
interactions are faster than the rate of expansion of the Universe, making $\tau$-leptons decouple from the flavour coherent lepton state. Hence, it is essential 
to study leptogenesis in a two-flavoured regime. Below $T\sim10^{9}$ GeV, the $\mu$-leptons decouple and completely break down the flavour coherence of the leptons. 
Hence, the study of leptogenesis requires three-flavour consideration. So we have a 
set of flavour-independent and flavour-specific Boltzmann equations given below, 
and we solve it numerically to obtain the unflavoured, and flavoured lepton 
asymmetries. As mentioned before, the Boltzmann equations \cite{Nardi:2006fx, Fong:2012buy, Gu:2006wj, Ahn:2007mj} 
take care of the dynamics of particle abundancies, taking care of the production 
of CP asymmetries and washing out of the asymmetry due to the interplay between
decay and inverse decay processes. These equations are expressed in terms of 
particle asymmetries of species $x$, $Y_{x}=\frac{n_{x}}{s}$ where $n_{x}$ is
the number density and $s$ is the entropy density.   
 \begin{equation}
  \frac{dY_{N}}{dz}=-\frac{\gamma_{D}}{sHz}\left(\frac{Y_{N}}{Y^{eq}_{N}}-1\right)
  \label{eq:N1-number}
  \end{equation}
  \begin{equation}
  \frac{dY_{\Delta_{i}}}{dz}=-\frac{\gamma_{D}}{sHz}\left[\left(\frac{Y_{N}}
  {Y^{eq}_{N}}-1\right)\epsilon_{i}+K^{i}_{0}\sum_{j}\frac{1}{2}\left(C^{l}_{ij}
  +C^{H}_{j}\right)\frac{Y_{\Delta_{j}}}{Y^{eq}_{l}}\right]
  \label{eq:FlavBEs}
 \end{equation}
where $z=\frac{M}{T}$, the decay and inverse decay rate of the right handed neutrino ($N \rightarrow l \phi$), $\gamma_{D}=DsHzY^{eq}_{N}$, 
$D=Kz\frac{K_{1}(z)}{K_{2}(z)}$, $s$ is the entropy density, $K_1$ and
$K_2$ are modified Bessel functions,
$Y_{\Delta_{i}}$ being the i-flavoured lepton asymmetry, 
$Y^{eq}_{N}=\frac{45}{2\pi^{4}g_{*}}z^{2}K_{2}(z)$, and 
$Y^{eq}_{l}=\frac{15}{4\pi^{2}g_{*}}$. The decay or wash-out parameter,
\begin{equation}
 K\equiv \frac{\sum_\alpha \Gamma(N \rightarrow L_\alpha \phi)}
 {H(M)} = \frac{\tilde{m}}{m_{*}},
 \label{eqn-Kmm}
\end{equation}
 where $\tilde{m}$ is the effective neutrino mass $\tilde{m}\equiv \frac{(Y^{\dagger}Y)_{11}v^{2}}{M}$ and is proportional to the total decay rate of the right-handed neutrino. $H(M)$ is the Hubble parameter evaluated at a temperature $T = M$; $m_{*}\sim  10^{-3}$ eV is equilibrium neutrino mass. 
We have considered the case of strong wash-out $(\tilde{m}>m_{*})$ for our study. $K^{i}_{0}$ are the flavour projection operators where $i=a,\tau$ in two-flavour configuration, and $i=e,\mu,\tau$ in three-flavour configuration. It can be expressed as \cite{Abada:2006ea, Fong:2012buy, Nardi:2005hs, Blanchet:2006be, Davidson:2008bu, Datta:2023pav},
 \begin{equation}
  K^{i}_{0}=\frac{\left(Y^{*}\right)_{i1}\left(Y\right)_{i1}}{\left(Y^{\dagger}Y\right)_{11}}.
  \label{eqn-flvproji}
 \end{equation}
The importance of flavour projection operators is discussed further in the appendix (\ref{sect-flavproj}).

In the case of flavour-independent lepton asymmetry, Eq.(\ref{eq:FlavBEs}) 
can be written as
\begin{equation}
  \frac{dY_{\Delta}}{dz}=-\frac{\gamma_{D}}{sHz}\left[\left(\frac{Y_{N}}{Y^{eq}_{N}}-1\right)
  \epsilon+\frac{1}{2}\frac{Y_{\Delta}}{Y^{eq}_{l}}\right],
  \label{eq:unFlavBEs}
 \end{equation}
 where $Y_{\Delta}$ represents the unflavoured lepton asymmetry.
 
\begin{figure}[h!]
\centering
 \includegraphics[width=12cm]{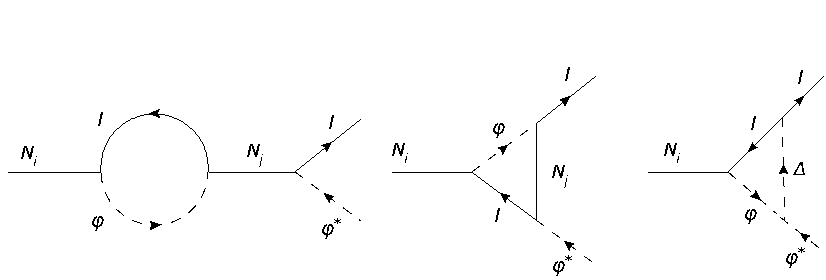}
 \caption{The one-loop Feynman diagrams of $N_{i}$ decay in a model 
 with $n$ number of right-handed neutrinos and one triplet, $\Delta$. In our case, we have one right-handed neutrino, so CP violation comes
 from the interference of tree-level diagram with the third diagram.}
 \label{fig-jaxo1}
\end{figure}

In this study, the lepton number violating decays 
of the heavy states introduced in the seesaw mechanism act as new sources
of CP violation. For our model, one right-handed neutrino $N$ and 
one heavier triplet scalar $\Delta$ are chosen as the two heavy states.
The lepton number violating effects produced by the heavier state get 
washed out due to the same produced by the lighter state. So, here the 
CP asymmetry is getting generated from the decay of the right-handed 
Majorana neutrino, with the assumption $M\ll M_{\Delta}$. The CP 
asymmetry can be calculated from the interaction of ordinary tree-level 
decay and the three diagrams shown in Fig.(\ref{fig-jaxo1}). The first two 
figures are self-energy and vertex diagrams mediated by one extra 
heavy right-handed neutrino. Nevertheless, there is only one 
right-handed neutrino in the model under consideration. So the 
third figure representing the vertex diagram mediated by the Higgs 
triplet scalar $\Delta$ contributes to the generation of CP asymmetry.
 In the temperature range $T>10^{12}$ GeV, the lepton flavours states
 can be approximated as a single or unflavoured state. In that temperature 
 range, we write the flavour-independent CP asymmetry parameter 
 as \cite{Gu:2006wj}
 \begin{equation}
  \epsilon^{\Delta}_{N}\simeq\frac{3}{16\pi}\frac{M}{v^{2}}
  \frac{\sum_{\alpha\beta}{\rm Im}\left[Y^{\dag}_{1\alpha}Y^{\dag}_{1\beta}
  (m^{(II)*}_{\nu})_{\alpha\beta}\right]}{\left(Y^{\dag}Y\right)_{11}}.
  \label{eq:eps1}
 \end{equation}
  Again, for temperature $T\ll10^{12}$ GeV, the lepton flavours become 
 distinguishable, and the flavour consideration becomes important. 
 Then the flavour-specific CP asymmetry parameter can be written 
 as \cite{Antusch:2004xy,Antusch:2007km}
 \begin{equation}
  \epsilon^{\Delta}_{N,\alpha}\simeq\frac{3}{16\pi}
  \frac{M}{v^{2}}\frac{\sum_{\beta}
  {\rm Im}\left[Y^{\dag}_{1\alpha}Y^{\dag}_{1\beta}
  (m^{(II)*}_{\nu})_{\alpha\beta}\right]}{\left(Y^{\dag}Y\right)_{11}}.
  \label{eq:eps2}
 \end{equation}

For simplicity, we write the unflavoured CP asymmetry as $\epsilon$ and the flavour
specific CP asymmetry as $\epsilon_{i}$, where $i=e$, $\mu$, $\tau$ or $a=e+\mu$.
From Eq.(\ref{eq:eps1}), we obtain the expression of the unflavoured CP asymmetry
parameter, using the neutrino mass matrix elements
\begin{equation}
  \epsilon=\frac{3}{16\pi}\frac{M}{v^{2}\left(1+r^{2}\right)}m_{0}\left(\hat{A_{\nu}}
  \sin\gamma_{\nu}+2r\hat{B_{\nu}}\sin\beta_{\nu}\right).
  \label{eq:CPuf}
 \end{equation}
From Eq.(\ref{eq:eps2}), we can further obtain the two-flavoured
CP asymmetries $\epsilon_{a}$, $\left(a=e+\mu\right)$  and $\epsilon_{\tau}$, 
\begin{equation*}
 \epsilon_{a}=\frac{3}{16\pi}\frac{M}{v^{2}\left(1+r^{2}\right)}rm_{0}\hat{B_{\nu}}\sin\beta_{\nu},
\end{equation*}
and 
\begin{equation}
 \epsilon_{\tau}=\frac{3}{16\pi}\frac{M}{v^{2}\left(1+r^{2}\right)}m_{0}
 \left(r\hat{B_{\nu}}\sin\beta_{\nu}+\hat{A_{\nu}}\sin\gamma_{\nu}\right),
 \label{eq:CP2fa}
\end{equation}
which is relevant in the temperature range $T>10^{9}$ GeV.
 
The CP asymmetry parameters are to be determined from 
the model to further investigate different lepton asymmetries
through a set of Boltzmann equations. 
In the temperature range $10^{11}\rm{GeV}
\lesssim T\lesssim 10^{12}\rm{GeV}$, under two-flavoured leptogenesis regime,
\begin{equation}
    C^{H}=\frac{1}{230}\left(41,56\right),
\end{equation}
\begin{equation}
 C^{l}=\frac{1}{460}
 \begin{pmatrix}
  196 & -24\\
  -9 & 156
 \end{pmatrix}.
\end{equation}
Finally, we can estimate the baryon asymmetry after solving the suitable set of Boltzmann equations numerically and obtain
the lepton asymmetry.
In the case of unflavoured leptogenesis, we use the expression,
\begin{equation}
 \eta_{B}=-7.04\times Y_{B}, \quad Y_{B}=-1.38\times10^{-3}\epsilon \eta,
 \label{eq:uetaB}
\end{equation}
to calculate baryon asymmetry \cite{Nardi:2006fx}, where 
\begin{equation}
 \eta=\frac{Y_{\Delta}\left(z\gg 1\right)/\epsilon}{Y^{eq}_{N}\left(0\right)},
\end{equation}
is known as the efficiency factor \cite{Fong:2013wr}. 
In the case of flavoured leptogenesis, the expression of baryon asymmetry given in 
Eq.(\ref{eq:uetaB}) is replaced by \cite{Nardi:2006fx}
\begin{equation}
 \eta_{B}=-7.04\times Y_{B}, \quad Y_{B}=-1.38\times10^{-3}
 \sum_{i} Y_{\Delta_{i}}\left(z\gg1\right).
 \label{eq:etaB}
\end{equation}
In the next subsection we make a quantitative analysis of the BAU in the model.
\subsection{Baryon asymmetry determination: Result}
\label{sec:result-bau}
 In this section, we make quantitative analysis of baryon asymmetry by calculating
the CP asymmetry and solving the set of Boltzmann equations, both for 
unflavoured and flavoured leptogenesis. The CP asymmetries are calculated
by using equations (\ref{eq:CPuf})- (\ref{eq:CP2fa}) for unflavoured and flavoured leptogenesis respectively. Similarly the corresponding set of 
Boltzmann equations (\ref{eq:N1-number}), (\ref{eq:FlavBEs}) and (\ref{eq:unFlavBEs})
are taken into account to calculate the final BAU using equations (\ref{eq:uetaB}) and 
(\ref{eq:etaB}). It is known that the generation of BAU can be enhanced by taking
flavour effects into account over unflavoured leptogenesis. Here, we show that 
Fritzsch-type textures can be used to verify the above characteristic of leptogenesis mechanism.
So an integrated scenario of generation of baryon asymmetry and experimentally compatible
values of neutrino mixing parameters can be achieved in this framework.

Although unflavoured leptogenesis is viable above temperature  $T \gtrsim 10^{12}$ GeV, we make
comparison between unflavoured and flavoured leptogenesis by bringing down the leptogenesis scale to $10^{10} - 10^{11}$ GeV, suitable for studying two-flavoured leptogenesis.
 In order to study the cases of unflavoured and two-flavoured leptogenesis, we choose different  values of $M$ in the range $4\times10^{10}$ GeV $\le M \le 5 \times 10^{11}$ GeV, and four benchmark sets are formed corresponding to those $M$ values. For each benchmark set, the values of neutrino mass eigenvalue $m_{1}$, phases $\gamma_{\nu}$, $\beta_{\nu}$, and $y_{0}$ are made to be fixed for calculating the CP asymmetry. The values are consistent with the
 oscillation data. Using the chosen values for these parameters the effective neutrino mass $\tilde{m}=\frac{(Y^{\dagger}Y)_{11}v^{2}}{M}=\frac{y^{2}_{0}(1+r^{2})}{M}$,  neutrino mass matrix elements $\hat{A_{\nu}}$, $\hat{B_{\nu}}$, $\hat{C_{\nu}}$ (using expressions shown in equations (\ref{eqn-Ahat}), (\ref{eqn-Bhat}) and (\ref{eqn-Chat})) are calculated for each set and shown in table (\ref{table-ABC}). For each set, the values of the parameters are chosen so that the neutrino mass eigenvalues follow normal
 hierarchy and the sum of neutrino masses, $\Sigma$ is in 
 agreement with its observational value. 
   
   \begin{table}[h!]
 \centering 
 \begin{tabular}{ |p{1.2cm}||p{1.5cm}||p{1.1cm}||p{1.5cm}||p{1.2cm}||p{1.2cm}||p{1.2cm}||p{1.2cm}||p{1.1cm}||p{1.1cm}|}
  \hline 
 Set no. & $M({\rm GeV})$&$y_{0}$&$\tilde{m}$(eV)&$\gamma_{\nu}$&$\beta_{\nu}$&$\hat{A}_{\nu}$ (eV)
 &$\hat{B}_{\nu}$ (eV)&$\hat{C}_{\nu}$ (eV)&$\Sigma$ (eV)\\
 \hline
 I&$4\times 10^{11}$&$0.0201$&$0.043379$&$57.32^{\circ}$&$343.95^{\circ}$
 &$-0.317$&$-0.177$&$0.461$&$0.0614$\\
 \hline
  II&$2\times 10^{11}$&$0.0201$&$0.074759$&$171.97^{\circ}$&$171.97^{\circ}$
  &$1.507$&$0.733$&$0.231$&$0.0684$\\
 \hline
 III&$8\times 10^{10}$&$0.0049$&$0.022743$ & $343.95^{\circ}$ &$57.32^{\circ}$ 
 & $2.172$ & $0.692$ & $1.528$ & $0.0644$ \\
 \hline
 IV&$4\times 10^{10}$&$0.0048$&$0.031625$&$57.32^{\circ}$&$114.65^{\circ}$
 &$0.827$&$0.669$&$0.796$&$0.0645$\\
 \hline
\end{tabular}
\caption{Neutrino mass matrix elements for different values 
of right-handed neutrino mass $M$. These values are used to estimate the CP asymmetry and wash-out parameters.}
\label{table-ABC}
\end{table}
For the hierarchical mass spectrum of light neutrinos, the
renormalization group running between low energy and the high energy seesaw scale has a nominal impact on the neutrino parameters, except from an overall scaling of the light neutrino masses \cite{Chankowski:1993tx,Casas:1999tg,Chankowski:2001mx}. The effect of scaling can be taken care of by multiplying by a factor of $1.2$ in the low energy values of the parameters \cite{Antusch:2003kp}.

The purpose of setting up these benchmark points is to determine and compare the production baryon asymmetry via unflavoured and two-flavoured leptogenesis. Hence, wash-out parameter $K=\frac{\tilde{m}}{m_{*}}=\frac{y^{2}_{0}(1+r^{2})v^{2}}{M m_{*}}$ and the CP asymmetries $\epsilon$, $\epsilon_{a}$, and $\epsilon_{\tau}$ are calculated corresponding to different $M$ values. For each set, baryon asymmetry is calculated via both unflavoured and two-flavoured leptogenesis. For the purpose of having a comparative study, we have kept the values of washout parameter $K$ and the CP asymmetry parameter $\epsilon(\epsilon_{i})$ consistent for the study of unflavoured and two-flavoured leptogenesis. The consistency is ensured by the relation $\epsilon=\epsilon_{a}+\epsilon_{\tau}$. In case of two-flavoured leptogenesis, the flavour projection operators $K^{i}_{0}$, expressed in Eq.(\ref{eqn-flvproji}), are calculated as 
  \begin{equation}
   K^{a}_{0}=\frac{(Y^{*})_{e1}(Y)_{e1}}{(Y^{\dagger}Y)_{11}}+\frac{(Y^{*})_{\mu1}(Y)_{\mu1}}{(Y^{\dagger}Y)_{11}}=\frac{r^{2}}{1+r^{2}},
  \end{equation}
 and 
\begin{equation}
   K^{\tau}_{0}=\frac{(Y^{*})_{\tau1}(Y)_{\tau1}}{(Y^{\dagger}Y)_{11}}=\frac{1}{1+r^{2}},
  \end{equation}
for each benchmark set.

The flavour independent, as well as flavour-specific CP asymmetry parameters arising from the model, 
mentioned in equations (\ref{eq:CPuf} - \ref{eq:CP2fa}), respectively, are functions of the 
neutrino mass eigenvalues $m_{1}$, $m_{2}$, $m_{3}$ through the neutrino mass matrix elements $\hat{A_\nu}$,
$\hat{B_\nu}$, $\hat{C_\nu}$. The mass eigenvalues $m_{2}$, $m_{3}$ and thereby the elements $\hat{A_\nu}$, $\hat{B_\nu}$,
$\hat{C_\nu}$ are calculated using the equations (\ref{eqn-dm213l}-
\ref{eqn-Chat}) for different values of $m_1$. The allowed ranges of $x_\nu$ and $y_\nu$, given in Eq.(\ref{eqn-xyrnge}), are used to estimate the CP asymmetry parameters. 

The suitable set of Boltzmann equations are solved numerically and they are shown in the appendix(\ref{app:BE-sol}).
The final lepton asymmetries thus obtained are used to calculate the final baryon asymmetries.
The final results corresponding to different sets are enlisted in table (\ref{table-setunf}) and table (\ref{table-settwof}), for unflavoured and two-flavoured leptogenesis, respectively. The corresponding plots are shown in Fig.(\ref{fig-fbl1}) in the appendix (\ref{app:BE-sol}). The baryon asymmetries $|\eta_{B}|$ are compared in table (\ref{table-baryonasymtable2}) to realize the enhancement in the result after considering leptogenesis in an appropriate flavoured regime. 
\begin{table}[h!]
 \centering 
 \begin{tabular}{ |p{1.2cm}||p{2.0cm}||p{1.8cm}||p{2.5cm}||p{2.2cm}| }
 \hline
 Set no.&$M({\rm GeV})$&K&$\epsilon$&$|\eta_{B}|/10^{-10}$\\
 \hline
 I&$4\times 10^{11}$&$43.379$&$-3.45\times 10^{-6}$&$1.78$\\
 \hline
 II&$2\times 10^{11}$&$74.759$&$6.12\times 10^{-6}$&$1.67$\\
 \hline
 III&$8\times 10^{10}$&$22.473$&$4.73\times 10^{-7}$&$0.54$\\
 \hline
 IV&$4\times 10^{10}$&$31.625$&$1.42\times 10^{-6}$&$1.07$\\
 \hline
\end{tabular}
\caption{Baryon asymmetries from unflavoured leptogenesis for different values of $M$ and $K$.}  
\label{table-setunf}
\end{table}

\begin{table}[h!]
 \centering 
 \begin{tabular}{ |p{1.2cm}||p{1.6cm}||p{1.3cm}||p{2.1cm}||p{2.3cm}||p{1.6cm}||p{1.6cm}||p{1.9cm}| }
 \hline
 Set no.&$M({\rm GeV})$&K&$\epsilon_{a}$&$\epsilon_{\tau}$&$K^{a}_{0}$&$K^{\tau}_{0}$&$|\eta_{B}|/10^{-10}$\\
\hline
I&$4\times 10^{11}$&$43.379$&$5.43\times 10^{-7}$&$-3.99\times 10^{-6}$&$0.294911$&$0.705088$&$2.02$\\
 \hline
 II&$2\times 10^{11}$&$74.759$&$9.61\times 10^{-7}$&$5.16\times 10^{-6}$&$0.181698$&$0.818301$&$5.10$\\
 \hline
 III&$8\times 10^{10}$&$22.473$&$4.19\times 10^{-7}$&$5.38\times 10^{-8}$&$0.588344$&$0.411655$&$1.56$\\
 \hline
 IV&$4\times 10^{10}$&$31.625$&$4.28\times 10^{-7}$&$9.92\times 10^{-7}$&$0.429678$&$0.570321$&$3.26$\\
 \hline
\end{tabular}
\caption{Baryon asymmetries from two-flavoured leptogenesis for different values of $M$, $K$, $\epsilon_{i}$ and $K^{i}_{0}$.}  
\label{table-settwof}
\end{table}

\begin{figure}[h!]
    \centering
\includegraphics[width=0.7\textwidth]{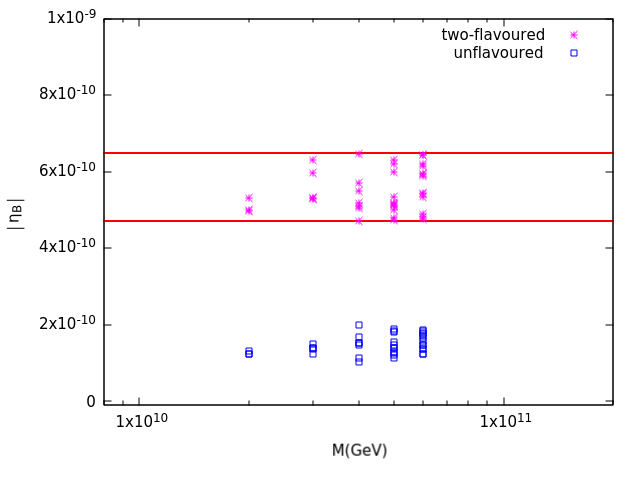}
  \caption{Comparing the Baryon asymmetry results coming from unflavoured and two-flavoured leptogenesis regimes. The blue points correspond to baryon asymmetries via unflavoured leptogenesis. On the other hand, the magenta points correspond to baryon asymmetries via two-flavoured leptogenesis. The region between the two red lines signifies the experimentally obtained range of baryon asymmetry, i.e. $[4.7-6.5]\times 10^{-10}$. }
  \medskip\small\raggedright 
  \label{fig-compare}
  \end{figure}

\begin{table}[h]
 \centering 
 \begin{tabular}{ |p{2.0cm}||p{6.0cm}||p{6.0cm}| }
  \hline
 $M({\rm GeV})$&$|\eta_{B}|$ from unflavoured leptogenesis&$|\eta_{B}|$ from two-flavoured leptogenesis\\
 \hline   
 $4\times 10^{11}$&$1.78\times10^{-10}$ [Set-I, \ref{fig-Ia}]&$2.02\times10^{-10}$ [Set-I, \ref{fig-Ib}]\\
 $2\times 10^{11}$
  &$1.67\times10^{-10}$ [Set-II, \ref{fig-IIa}]&$5.10\times10^{-10}$ [Set-II, \ref{fig-IIb}]\\
$8\times 10^{10}$
  &$5.38\times10^{-11}$ [Set-III, \ref{fig-IIIa}]&$1.56\times10^{-10}$ [Set-III, \ref{fig-IIIb}]\\
  $4\times 10^{10}$
  &$1.07\times10^{-10}$ [Set-IV, \ref{fig-IVa}]&$3.26\times10^{-10}$ [Set-IV, \ref{fig-IVb}]\\
 \hline
\end{tabular}
\caption{Baryon asymmetries from unflavoured and two-flavoured leptogenesis for different values of $M$, and 
$\epsilon(\epsilon_{i})$. We have shown a comparative study on how the flavour effects enhance 
the baryon asymmetry. The different sets in the table correspond to the corresponding plots, where the set of Boltzmann equations are numerically solved and lepton asymmetries are shown.}  
\label{table-baryonasymtable2}
\end{table}
In addition to using the benchmark points, to observe the importance of the flavour effects in producing baryon asymmetry through unflavoured and two-flavoured leptogenesis, we numerically vary the right-handed neutrino mass $M$ values in the range $[2-6]\times10^{10}$ GeV, for different values of $K$ and $\epsilon$ ($K$, $K^{i}_{0}$ and $\epsilon_{i}$ for flavoured leptogenesis) and show the results in Fig.(\ref{fig-compare}). In Fig.(\ref{fig-compare}), baryon asymmetries are plotted against $M$. After eliminating the results showing over-production of baryon asymmetry $|\eta_{B}|$, it shows that within the experimental bound $\left(4.7-6.5\right)\times 10^{-10}$, the results coming from two-flavoured leptogenesis are enhanced over the results coming from unflavoured leptogenesis.

\section{Relating CP violation in low and high energy phenomena}
\label{sec:cp-low-high}
Baryogenesis through leptogenesis can easily be implemented in the seesaw models.    
CP violation in lepton sector which are measurable in low-energy experiments like neutrino factories can have profound consequences in high-energy phenomena like leptogenesis. It is believed that the CP violation in the two sectors is in general not related to each other. The difficulty in setting
up the relation is due to lack of low-energy data available to 
quantify the parameters of the seesaw model. There are interesting studies 
\cite{Branco:2001pq} which show that it may be possible in specific Grand Unification
inspired models up to size and sign of the observed BAU to CP violation at 
low energies. The link between CP violations in leptogenesis and low-energy observable like neutrino less double beta decay, lepton flavour violation, Jarlskog invariant has been studied in 
\cite{Branco:2002kt,Branco:2009mb,Ellis:2002xg,Endoh:2002dk,Frampton:2002qc,Hagedorn:2016lva,Hagedorn:2017wjy,Joshipura:2001ui,Karmakar:2015jza,Li:2022bqy,Moffat:2018smo,Rebelo:2002wj}. 
In our work, we encounter a common origin of CP violation in low-energy neutrino experiments in terms of $J_{\rm CP}$ and 
in high-energy sector in terms of CP asymmetry parameter, required for leptogenesis. It can be
explained in a geometrical interpretation of CP violation with Majorana neutrinos. 
   
Analogous to the CKM matrix in the quark sector, in the lepton sector,  six unitarity triangles can be formed known as leptonic
unitarity triangles, from the orthogonality of the rows and columns of the $3\times 3$ PMNS matrix. These triangles are analogous to the quark unitarity triangles used for studying various manifestations of CP violation. However, in the case of Majorana neutrinos, there is an important difference which will be discussed here. The unitarity condition of $V$ is given by,
\begin{equation}
  V^{\dagger}V=VV^{\dagger}=1.
 \end{equation}
Under the rephasing transformation of lepton fields $ L_\alpha
 \rightarrow e^{i\phi_\alpha} L_\alpha$, the matrix $V$ transforms as $V_{\alpha i}\rightarrow V'_{\alpha i}=e^{i\phi_{\alpha}}V_{\alpha i}$. The vector $V_{\alpha i} V^{*}_{\beta i} \rightarrow e^{i(\phi_{\alpha} -\phi_\beta) } V_{\alpha i} V^{*}_{\beta i}$ rotates in the complex plane, whereas the vector $V_{\alpha i} V^{*}_{\alpha j}$ remains invariant. Based on this observation, the unitarity triangles are classified into Dirac triangles and Majorana triangles which are discussed below. 
 \subsection{Dirac Unitarity Triangles}
  From the orthogonality of rows of the mixng matrix $V$ , 
 \begin{equation}
  \Sigma V_{\alpha i} V^{*}_{\beta i}=0 \quad (\alpha\neq \beta),
  \label{eqn-DirU}
 \end{equation}
 we obtain the expressions of three Dirac triangles,
 \begin{equation}
  T_{e\mu}: V_{e1}V^{*}_{\mu1}+V_{e2}V^{*}_{\mu2}+V_{e3}V^{*}_{\mu3}=0,
  \label{eqn-Temu}
 \end{equation}
 \begin{equation}
  T_{e\tau}: V_{e1}V^{*}_{\tau1}+V_{e2}V^{*}_{\tau2}+V_{e3}V^{*}_{\tau3}=0,
  \label{eqn-Tetau}
 \end{equation}
 \begin{equation}
  T_{\mu\tau}: V_{\mu1}V^{*}_{\tau1}+V_{\mu2}V^{*}_{\tau2}+V_{\mu3}V^{*}_{\tau3}=0.
  \label{eqn-Tmutau}
 \end{equation}
The orientation of the Dirac triangles has no physical significance since under the rephasing of the charged-lepton fields these triangles exhibit rotation in the complex plane. The Dirac
triangles share a common area $A= \frac{1}{2}J_{CP}$ and the vanishing area of the Dirac triangles indicates vanishing Jarlskog Invariant $J_{CP}=0$ but it does not guarantee the conservation of CP symmetry. It only indicates that the Dirac CP phase is zero, but the two Majorana phases can still violate CP. Thus Dirac triangles fail to completely describe CP violation \cite{Aguilar-Saavedra:2000jom}. 
      
\begin{figure}[h!]
\centering
\includegraphics[width=0.45\textwidth,trim={0 0.1cm 0 0},clip]{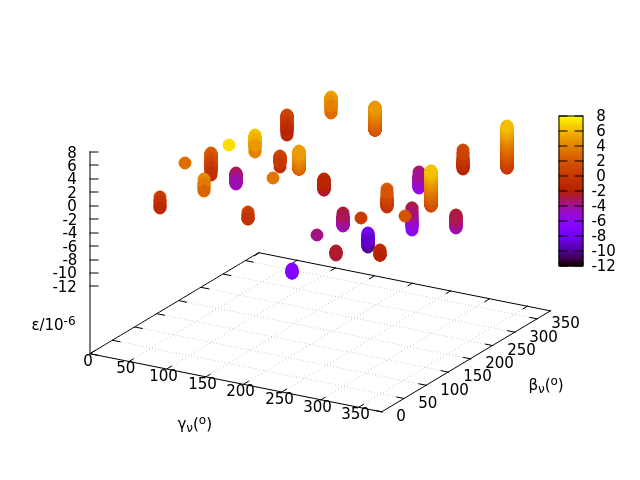}
\hspace{3ex}
\includegraphics[width=0.472\textwidth,trim={0 0 0 0.08cm},clip]{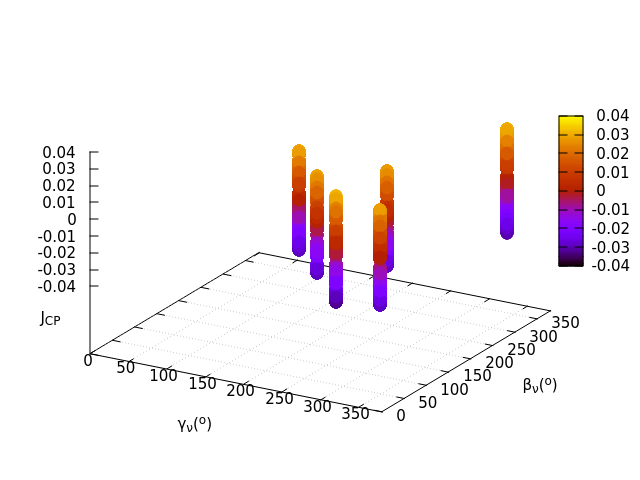}
\caption{ 3D scattered plots of $\epsilon$ and $J_{CP}$ as function of
$\beta_\nu$ and $\gamma_\nu$: The figure on the left shows the high-energy CP violation measured by CP asymmetry $\epsilon$ as a function of phases $\gamma_{\nu}$ and $\beta_{\nu}$ in a unit of $10^{-6}$, and the figure in right shows the low-energy CP violation measured by Jarlskog invariant $J_{CP}$ as a function of $\gamma_{\nu}$ and $\beta_{\nu}$. These plots show the common connection between low- and high-energy CP violation through the two phases $(\gamma_{\nu}, \beta_{\nu})$ appearing in the neutrino mass matrix. }
\label{fig:phases}
\end{figure}

Nevertheless, the quantity $J_{\rm CP}$ can be determined through
\begin{equation}
 J_{CP}=Im\left(V_{11}V_{22}V^{*}_{12}V^{*}_{21}\right),
 \label{eq:Jcp}
\end{equation}
by using the explicit form of $V$. In this case, it is not possible to
get a compact form of $J_{\rm CP}$, we make an approximate 
analytical study in appendix(\ref{sec:jcp-eps}), which shows that 
it depends on the phases $\beta_\nu$ and $\gamma_\nu$. Also, it can be
observed from equations (\ref{eq:CPuf}) and (\ref{eq:CP2fa}), the same
phases appear in the CP asymmetry parameter. We numerically calculate
the values of $J_{\rm CP}$ and $\epsilon$ as shown in Fig.(\ref{fig:phases}). 
In order to calculate $\epsilon$ the value of $m_{1}$ is chosen to be $0.021$ eV, and $m_{2}$ and $m_{3}$ are further calculated following the relation given in Eq.(\ref{eqn-xyn}), using the allowed parameter space for $[x_{\nu},y_{\nu}]$ given in Eq.(\ref{eqn-xyrnge}).
The allowed neutrino mass eigenvalues $m_{1}$, $m_{2}$, $m_{3}$ are found by imposing the bound on total neutrino mass $\Sigma=m_{1}+m_{2}+m_{3}<0.12$ eV.
The parameter $y_{0}$ is varied in the range $[0.0101:0.0201]$. The right-handed neutrino mass $M$ is chosen to be $8\times 10^{10}$ GeV. The phases $[\gamma_{\nu},\beta_{\nu}]$ are both varied in the range $[0:2\pi]$. The maximum order of the obtained CP asymmetry is found to be $|\epsilon|\sim10^{-5}$. 

On the other hand, in the expression of $J_{CP}$ in Eq.(\ref{eq:Jcp}), the mixing matrix $V$ is a function of $U_{\nu}$, $U_{l}$ and phases $(\theta,\phi,\psi)$ as it can be seen in Eq.(\ref{eqn-vpq}).
Here $U_{\nu}$ and $U_{l}$ are given in appendix (\ref{app-1}) and (\ref{app-2}), respectively. The elements of $U_{l}(U_{\nu})$ are functions of charged (neutral) lepton mass ratios $[x_{l}(x_{\nu}),y_{l}(y_{\nu})]$. The charged lepton mass ratios are determined, as given in Eq.(\ref{eqn-xyl}).
Also keeping in mind the factorization of lepton mass matrices given
in Eq.(\ref{eqn-mPmP}), can be written, when the
the condition imposed on elements of the neutrino mass matrix given in Eq.(\ref{eqn-omnu}) was ${\rm arg} \left(1+\hat{A_{\nu}} e^{i\gamma_\nu}\right) = 2 {\arg}
\left(r+ \hat{B_{\nu}}e^{i\beta_\nu}\right )$.  Using this condition one can see $\psi_{\nu}$ is related to the phases $[\gamma_{\nu},\beta_{\nu}]$ as,
 \begin{equation}
  \frac{\hat{A_{\nu}}\sin\gamma_{\nu}}{1+\hat{A_{\nu}}\cos\gamma_{\nu}}=\tan 2\psi_{\nu}, \quad
   \frac{\hat{B_{\nu}}\sin\beta_{\nu}}{r+\hat{B_{\nu}}\cos\beta_{\nu}}=\tan\psi_{\nu}.
  \label{eqn-Tpsi}
 \end{equation}
Therefore, it is understood that $J_{CP}$ is a function of the phases $[\gamma_{\nu},\beta_{\nu}]$ through $\psi_{\nu}$. 
Since, $\hat{A_{\nu}}$, $\hat{B_{\nu}}$, $\hat{C_{\nu}}$ and $r$ depend on $M$, $m_{1}$, $m_{2}$, $m_{3}$ and $y_{0}$, we vary $m_{1}\sim[0.001-0.05]$eV and $y_{0}\sim[0.0001-1]$ and obtained $m_{2}$, $m_{3}$ using Eq.(\ref{eqn-xyn}) within the allowed parameter space for $[x_{\nu},y_{\nu}]$ mentioned in Eq.(\ref{eqn-xyrnge}). We also varied $\theta$, $\phi$, $\psi_{l}$ and $[\gamma_{\nu},\beta_{\nu}]$ in the range $[0:2\pi]$, and determined the allowed range for $\psi_{\nu}$ and thereby for $[\gamma_{\nu},\beta_{\nu}]$, which satisfy the conditions given in Eq.(\ref{eqn-Tpsi}) for obtaining a scattered plot of $J_{CP}$ as a function of $[\gamma_{\nu},\beta_{\nu}]$. The maximum low energy CP violation through $|J_{CP}|$ for $M=8\times 10^{10}$ GeV is found to be $\sim 0.032$. 
\subsection{Majorana Unitarity Triangles}
From the orthogonality of columns of the mixing matrix, $V$,
 \begin{equation}
  \Sigma V_{\alpha i} V^{*}_{\alpha j}=0 \quad (i\neq j),
  \label{eqn-MajU}
 \end{equation}
 we obtain the expressions of three Majorana triangles,
 \begin{equation}
   T_{12}: V_{e1}V^{*}_{e2}+V_{\mu1}V^{*}_{\mu2}+V_{\tau1}V^{*}_{\tau2}=0,
   \label{eqn-T12}
 \end{equation}
 \begin{equation}
   T_{13}: V_{e1}V^{*}_{e3}+V_{\mu1}V^{*}_{\mu3}+V_{\tau1}V^{*}_{\tau3}=0,
   \label{eqn-T13}
 \end{equation}
 \begin{equation}
   T_{23}: V_{e2}V^{*}_{e3}+V_{\mu2}V^{*}_{\mu3}+V_{\tau2}V^{*}_{\tau3}=0.
   \label{eqn-T23}
 \end{equation}
 Since the Majorana triangles remain invariant under the rephasing, the orientation of the Majorana triangles has physical significance.
These Majorana triangles provide the necessary and sufficient conditions for CP conservation in the lepton sector. The absence of CP violation is guaranteed by
\begin{enumerate}
\item Vanishing of the common area $A = \frac{1}{2}J_{CP}$ of the Majorana triangles.
\item Orientation of all Majorana triangles along the direction of the real or imaginary axes.
\end{enumerate}
The first condition implies that the three triangles collapse into lines in the complex plane and the vanishing of the Dirac phase. The second condition implies that the Majorana phases do not violate CP. Hence the three Majorana triangles are capable to provide a complete description of CP violation, unlike the Dirac triangles. If each one of the sides of the Majorana triangles is not parallel to one of the axes, then it is a signal for CP non-conservation, contrarily to the Dirac triangle case where only a nonzero area signifies CP violation \cite{Aguilar-Saavedra:2000jom}.

\begin{figure}[h!]
\centering
\includegraphics[width=0.45\textwidth,trim={0 0.1cm 0 0},clip]{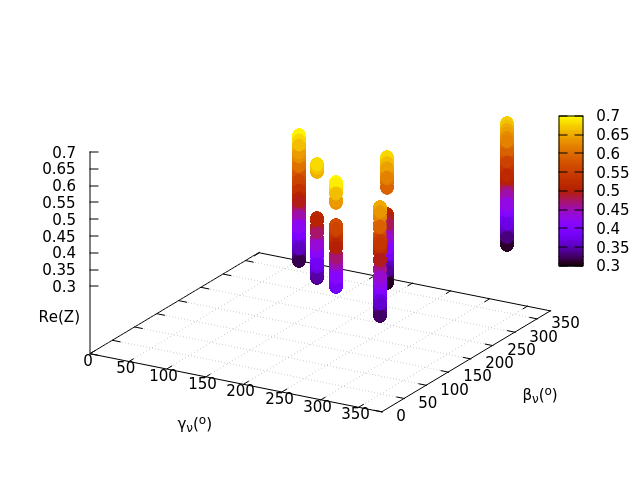}
\hspace{3ex}
\includegraphics[width=0.472\textwidth,trim={0 0 0 0.08cm},clip]{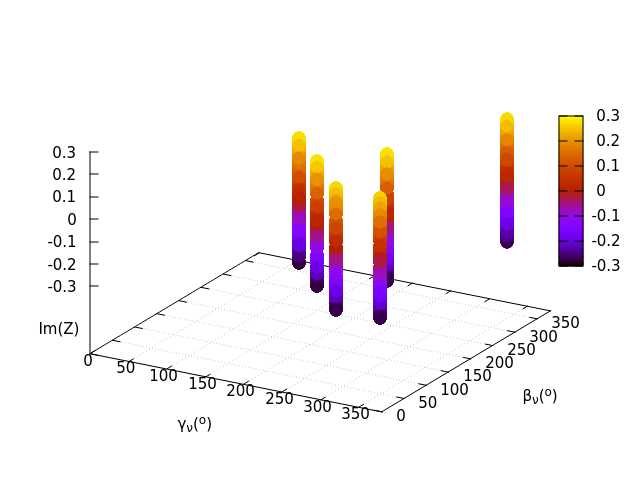}
\caption{ 3D scattered plots of ${\rm Re}(Z)(\gamma_{\nu},\beta_{\nu})$ and ${\rm Im}(Z)(\gamma_{\nu},\beta_{\nu})$: The plots show the dependence of the $Z$-parameter on the CP violating phases arising in the neutrino mass matrix, for the case of Majorana triangle-$T_{13}$ (Eq.(\ref{eqn-T13})). The figure in left(right) shows ${\rm Re}(Z)({\rm Im}(Z))$ as functions of the two phases $(\gamma_{\nu}, \beta_{\nu})$ appearing in neutrino mass matrix.}
\medskip\small\raggedright   
\label{fig-Zplot}
\end{figure}

The low energy CP violation can be obtained through the lepton unitarity triangles as discussed earlier. We consider the Majorana triangle, $T_{13}$ as expressed in Eq.(\ref{eqn-T13}), and define a parameter $Z$ as a side of the Majorana triangle-$T13$ after resizing it so that the base of the triangle becomes of unit length. In analogy with the quark
sector, the triangle corresponding to the unitary conditions in the first and third columns with
proper rescaling is shown in a figure in NuFIT 5.0 \cite{Esteban:2020cvm}. The figure indicates the absence of CP violation
would imply a flat triangle i.e., ${\rm Im}(Z) = 0$, where 
 \begin{equation}
  Z=-\frac{V_{e1}V^{*}_{e3}}{V_{\mu1}V^{*}_{\mu3}}={\rm Re}(Z)+i~ {\rm Im}(Z).
 \end{equation}
 As long as the area of the triangle is non-zero and all Majorana triangles are oriented along neither real nor imaginary axes, the CP symmetry will be violated. In order to understand the dependence of the low energy CP violation through the $Z$-parameter, on the CP violating phases arising in the neutrino mass matrix, the real and imaginary counterpart of $Z$-parameter is plotted as functions of the two phases $(\gamma_{\nu}, \beta_{\nu})$ as depicted in Fig.(\ref{fig-Zplot}). 
Here, the mixing matrix $V$ is found from the Eq.(\ref{eqn-vpq}).
From the Eq.(\ref{eqn-Tpsi}), it can be seen that $Z$-parameter is a function of the phases $[\gamma_{\nu},\beta_{\nu}]$ through $\psi_{\nu}$. 

For right-handed neutrino mass $M=8\times 10^{10}$ GeV, we varied $m_{1}\sim[0.001-0.05]$ eV and $y_{0}\sim[0.0001-1]$ and obtained $m_{2}$, $m_{3}$ through Eq.(\ref{eqn-xyn}) within the allowed parameter space for $[x_{\nu},y_{\nu}]$ given in Eq.(\ref{eqn-xyrnge}). We also varied $\theta$, $\phi$, $\psi_{l}$ and $[\gamma_{\nu},\beta_{\nu}]$ in the range $[0:2\pi]$, and determined the allowed range for $\psi_{\nu}$ and thereby for $[\gamma_{\nu},\beta_{\nu}]$ which satisfy the conditions given in Eq.(\ref{eqn-Tpsi}) for obtaining a scattered plot of ${\rm Re}(Z)$ and ${\rm Im}(Z)$ as a function of $[\gamma_{\nu},\beta_{\nu}]$. 

\section{Conclusion}
   \label{sec-Conclusion}
 In the minimal extension of the SM with right-handed neutrinos and scalar triplet,
 baryogenesis can be achieved through leptogenesis from the CP violating decay of
 either the lightest right-handed neutrino or triplet scalar. We have studied a minimal type-II seesaw model where the SM is extended with one right-handed neutrino and one triplet scalar. There are two mass scales
 involved in this case; the mass of the right-handed neutrino, $M$, and that
 of the triplet scalar, $M_\Delta$. Considering there are no heavy scalars in the theory
 we choose to work in the hierarchical mass $M \ll M_\Delta$.  In this case, for
 leptogenesis, the sources of CP violation can be found, that are mediated by the decay of the  
 heavy right-handed neutrino. In the absence of extra right-handed
 neutrinos, non-vanishing CP asymmetry is sourced from the interference of the tree-level and one-loop diagrams mediated by the Higgs triplet scalar, which is taken to be heavier than the 
 right-handed neutrino. By lowering the mass scale of the heavy right-handed neutrino, its  mass is 
 taken to be in the range $M \subset\left[10^{10},10^{11}\right]$ 
 GeV for sucessful baryogenesis via leptogenesis.
 In this mass range, we have studied the leptogenesis in a 
 two-flavoured regime. 
It is 
 seen that for $M \subset\left[10^{10},10^{11}\right]$ GeV, the obtained lepton asymmetries
 lead to baryon asymmetry within the desired range of 
 $\eta_{B}\sim\left(4.7-6.5\right)\times 10^{-10}$. We show 
 that incorporating appropriate flavour consideration, the results show an enhancement  in the baryon asymmetry as compared to unflavoured case. Thus the minimal type-II seesaw model with
 only one heavy right-handed neutrino and one heavier Higgs triplet
 scalar can provide a viable explanation for neutrino mass generation
 and baryon asymmetry of the Universe through leptogenesis. We show this feature
 by using the Fritzsch-type texture. 
 Using geometrical interpretation of low-energy CP violation we also show
 there is a common link between CP violation of both low- and high-energy regimes.
 These features can further be implemented in more predictive models like left-right symmetric models.
\appendix
\section{Diagonalizing neutrino mass matrix 
$m_{\nu}$ with two-zero texture}
\label{app-1}
The elements of the matrix $U_{\nu}$ are given in terms of the ratios $x_{\nu}$, $y_{\nu}$:
\begin{equation*}
 {U_\nu}_{11}=i\left[\frac{1}{\left(1+x_{\nu}\right)\left(1-x^{2}_{\nu}y^{2}_{\nu}\right)}\right]^{\frac{1}{2}},
\end{equation*}
\begin{equation*}
{U_\nu}_{12}=+\left[\frac{x_{\nu}\left(1-y_{\nu}-x_{\nu}y_{\nu}\right)}{\left(1+x_{\nu}\right)\left(1-y_{\nu}\right)\left(1-x_{\nu}y_{\nu}\right)}\right]^{\frac{1}{2}},
\end{equation*}
\begin{equation*}
 {U_\nu}_{13}=+\left[\frac{x^{2}_{\nu}y^{3}_{\nu}}{\left(1-y_{\nu}\right)\left(1-x^{2}_{\nu}y^{2}_{\nu}\right)}\right]^{\frac{1}{2}},
\end{equation*}
\begin{equation*}
{U_\nu}_{21}=-i\left[\frac{x_{\nu}}{\left(1+x_{\nu}\right)\left(1+x_{\nu}y_{\nu}\right)}\right]^{\frac{1}{2}},
\end{equation*}
\begin{equation*}
{U_\nu}_{22}=+\left[\frac{\left(1-y_{\nu}-x_{\nu}y_{\nu}\right)}{\left(1+x_{\nu}\right)\left(1-y_{\nu}\right)}\right]^{\frac{1}{2}},
\end{equation*}
\begin{equation*}
 {U_\nu}_{23}=+\left[\frac{x_{\nu}y_{\nu}}{\left(1-y_{\nu}\right)\left(1+x_{\nu}y_{\nu}\right)}\right]^{\frac{1}{2}},
\end{equation*}
\begin{equation*}
 {U_\nu}_{31}=+i\left[\frac{x^{2}_{\nu}y_{\nu}\left(1-y_{\nu}-x_{\nu}y_{\nu}\right)}{\left(1+x_{\nu}\right)\left(1-x^{2}_{\nu}y^{2}_{\nu}\right)}\right]^{\frac{1}{2}},
\end{equation*}
\begin{equation*}
 {U_\nu}_{32}=-\left[\frac{x_{\nu}y_{\nu}}{\left(1+x_{\nu}\right)\left(1-y_{\nu}\right)\left(1-x_{\nu}y_{\nu}\right)}\right]^{\frac{1}{2}},
\end{equation*}
\begin{equation*}
 {U_\nu}_{33}=+\left[\frac{1-y_{\nu}-x_{\nu}y_{\nu}}{\left(1-y_{\nu}\right)\left(1-x^{2}_{\nu}y^{2}_{\nu}\right)}\right]^{\frac{1}{2}}.
\end{equation*}

\section{Diagonalizing charged-lepton mass matrix 
$m_{l}$ with three-zero texture}
\label{app-2}
The elements of the matrix $U_{l}$ are given in terms of the ratios $x_{l}$, $y_{l}$:
\begin{equation*}
{U_l}_{11}=+\left[\frac{1-y_{l}}{\left(1+x_{l}\right)\left(1-x_{l}y_{l}\right)\left(1-y_{l}+x_{l}y_{l}\right)}\right]^{\frac{1}{2}},
\end{equation*}
\begin{equation*}
 {U_l}_{12}=-i\left[\frac{x_{l}\left(1+x_{l}y_{l}\right)}{\left(1+x_{l}\right)\left(1+y_{l}\right)\left(1-y_{l}+x_{l}y_{l}\right)}\right]^{\frac{1}{2}},
\end{equation*}
\begin{equation*}
 {U_l}_{13}=+\left[\frac{x_{l}y^{3}_{l}\left(1-x_{l}\right)}{\left(1-x_{l}y_{l}\right)\left(1+y_{l}\right)\left(1-y_{l}+x_{l}y_{l}\right)}\right]^{\frac{1}{2}},
\end{equation*}
\begin{equation*}
 {U_l}_{21}=+\left[\frac{x_{l}\left(1-y_{l}\right)}{\left(1+x_{l}\right)\left(1-x_{l}y_{l}\right)}\right]^{\frac{1}{2}},
\end{equation*}
\begin{equation*}
 {U_l}_{22}=+i\left[\frac{1+x_{l}y_{l}}{\left(1+x_{l}\right)\left(1+y_{l}\right)}\right]^{\frac{1}{2}},
\end{equation*}
\begin{equation*}
 {U_l}_{23}=+\left[\frac{y_{l}\left(1-x_{l}\right)}{\left(1-x_{l}y_{l}\right)\left(1+y_{l}\right)}\right]^{\frac{1}{2}},
\end{equation*}
\begin{equation*}
 {U_l}_{31}=-\left[\frac{x_{l}y_{l}\left(1-x_{l}\right)\left(1+x_{l}y_{l}\right)}{\left(1+x_{l}\right)\left(1-x_{l}y_{l}\right)\left(1-y_{l}+x_{l}y_{l}\right)}\right]^{\frac{1}{2}},
\end{equation*}
\begin{equation*}
 {U_l}_{32}=-i\left[\frac{y_{l}\left(1-x_{l}\right)\left(1-y_{l}\right)}{\left(1+x_{l}\right)\left(1+y_{l}\right)\left(1-y_{l}+x_{l}y_{l}\right)}\right]^{\frac{1}{2}},
\end{equation*}
\begin{equation*}
 {U_l}_{33}=+\left[\frac{\left(1-y_{l}\right)\left(1+x_{l}y_{l}\right)}{\left(1-x_{l}y_{l}\right)\left(1+x_{l}\right)\left(1-y_{l}+x_{l}y_{l}\right)}\right]^{\frac{1}{2}}.
\end{equation*}

\section{The importance of flavour projectors in leptogenesis}
\label{sect-flavproj}
Before deepening the elaborate execution of flavoured leptogenesis,
we will see how lepton flavours play a role in constructing CP asymmetry.
In general, in the case of leptogenesis, CP violation can be observed from 
$N_{1}$ decay in two cases:
 \begin{enumerate}
  \item \label{item-CP1} If the rate of production of 
  leptons and anti-leptons differ, expressed as
  \begin{equation*}
   \Gamma\neq\overline{\Gamma},
  \end{equation*}
where $\Gamma$ is the decay rate of the process $N_{1}\longrightarrow l+\phi^{\dagger}$, and $\overline{\Gamma}$ is the decay rate of the process $N_{1}\longrightarrow \overline{l'}+\phi$. The CP asymmetry can be expressed in terms of these decay rates, as
\begin{equation}
 \epsilon=\frac{\Gamma-\overline{\Gamma}}{\Gamma+\overline{\Gamma}}.
\end{equation}
\item \label{item-CP2} If the lepton flavour effects are included, 
it can affect the CP asymmetry construction in two ways,
\begin{enumerate}
\item It suppresses the washout, as the interaction of the leptons
with Higgs, during inverse decay, gets fragmented in terms of different
flavour states. In this context, we define a parameter called flavour
projector\cite{Nardi:2006fx, Blanchet:2006ch,Blanchet:2006be, Dev:2017trv}, 
\begin{equation*}
 K_{i}=\left|<l|l_{i}>\right|^{2}=\frac{\Gamma_{i}}{\Gamma}, \quad i=e,\mu,\tau
\end{equation*}
and 
\begin{equation*}
 \overline{K}_{i}=\left|<\overline{l'}\right|\overline{l}_{i}>|^{2}=\frac{\overline{\Gamma}_{i}}{\overline{\Gamma}}. 
\end{equation*}
Here, $\Gamma_{i}$ is the partial decay rate of the process $N_{1}\longrightarrow l_{i}+\phi^{\dagger}$, and $\overline{\Gamma}_{i}$ is the partial decay rate of the process $N_{1}\longrightarrow \overline{l}_{i}+\phi$. The concept of total decay rate $\Gamma=\sum_{i}\Gamma_{i}$, and $\overline{\Gamma}=\sum_{i}\overline{\Gamma}_{i}$ leads to the realisation that $\sum_{i} K_{i}=\sum_{i}\overline{K}_{i}=1$. Due to these flavour effects, we need to consider individually flavoured CP asymmetries as 
\begin{equation}
 \epsilon_{i}=\frac{\Gamma_{i}-\overline{\Gamma}_{i}}{\Gamma_{i}+\overline{\Gamma}_{i}}.
\end{equation}
\item If the state $|\overline{l'}>$ is not the CP conjugate state 
of the state $|l>$, which arises from misalignment in the flavour space due to loop-effects,
then it gives an additional source of CP violation.
To understand this effect, we introduce projector difference, as
\begin{equation*}
 \Delta K_{i}=K_{i}-\overline{K}_{i}.
\end{equation*}
The CP asymmetry can be further modified, incorporating $K_{i}=K^{i}_{0}+\frac{\Delta K_{i}}{2}$ and $\overline{K}_{i}=K^{i}_{0}-\frac{\Delta K_{i}}{2}$, as 
\begin{equation}
 \epsilon_{i}\sim\epsilon K^{i}_{0}+\frac{\Delta K_{i}}{2}.
 \label{eqn-CPK}
\end{equation}
Here, $K^{i}_{0}=\frac{K_{i}+\overline{K}_{i}}{2}$ is the tree level
contributions to the projections. 
\end{enumerate}
 \end{enumerate}
 In the Eq.(\ref{eqn-CPK}), the first term on the right-hand side,
 proportional to $\epsilon$, comes from the type of contribution 
 discussed in the first case. On the other hand, the second 
 term proportional to $\Delta K_{i}$ comes from the contribution 
 mentioned in the second case, as the term vanishes when we 
 have $|\overline{l'}\rangle$ as the CP conjugate state of $|l\rangle$. From 
 Eq.(\ref{eqn-CPK}), it can be easily shown that $\epsilon=\sum_{i} 
 \epsilon_{i}$. As we are particularly interested in a temperature 
 range where lepton flavour interaction becomes important, we have 
 incorporated flavour projectors, especially the tree level 
 contribution $K^{i}_{0}$ in the flavoured Boltzmann equations,
 which we have described in the next section. Flavour projectors 
 become important to segregate the flavour regimes of leptogenesis 
 in terms of lepton flavour alignment or non-alignment. In the 
 context of our chosen temperature range, the lepton flavour 
 non-alignment problem will be relevant, which refers to the 
 situation when no flavour state can be found to be perfectly 
 aligned with the states $|l\rangle$ and $|l'\rangle$ \cite{Nardi:2006fx}. 
 
 When only the $\tau$-Yukawa processes are in thermal equilibrium, 
 then the concept of flavour projectors suggests, 
 \begin{equation*}
  \sum_{i=a,\tau} K_{i}=K_{a}+K_{\tau}=1,
 \end{equation*}
and 
\begin{equation*}
  \sum_{i=a,\tau} \overline{K_{i}}=\overline{K}_{a}+\overline{K}_{\tau}=1,
 \end{equation*}
 where, $|l_{a}\rangle$ and $|\overline{l'}_{a}\rangle$ are two entangled states
 of flavours $e$ and $\mu$. Generally, this condition appears as a 
 two-flavoured leptogenesis scenario. On the other hand, when $\tau$ 
 and $\mu$-Yukawa processes are in thermal equilibrium, no entangled 
 flavour states can be formed. The thermal bath becomes populated with
 the CP-conjugate flavour states $|l_{i}\rangle$ and $|\overline{l}_{i}\rangle$ 
 (with $i=e$, $\mu$, $\tau$). Here arises typically the case of 
 three-flavoured or fully-flavoured leptogenesis. From the flavour 
 projector condition, we obtain,
 \begin{equation*}
  \sum_{i=e,\mu,\tau}K_{i}=K_{e}+K_{\mu}+K_{\tau}=1,
 \end{equation*}
and 
\begin{equation*}
  \sum_{i=e,\mu,\tau}\overline{K}_{i}=\overline{K}_{e}+\overline{K}_{\mu}+\overline{K}_{\tau}=1.
 \end{equation*}
\section{Low scale CP violation and Jarlskog invariant $J_{CP}$}
\label{sec:jcp-eps}
  In simplified form, 
 \begin{equation}
 J_{CP}\sim {\rm Im} \left[V_{13}\right],
 \label{eqn-imv}
 \end{equation}
 where $V_{13}$ can be obtained from Eq.(\ref{eqn-vpq}) as,
 \begin{equation}
  V_{13}={U_l}_{11}{U_\nu}^{*}_{13}e^{i\theta}+{U_l}_{21}{U_\nu}^{*}_{23}
  e^{i\phi}+{U_l}_{31}{U_\nu}^{*}_{33}e^{i\psi}. 
 \label{eqn-v13}
 \end{equation}
 From appendix(\ref{app-1}), it is clear that ${U_l}_{11}$, ${U_l}_{21}$, 
 ${U_l}_{31}$, ${U_\nu}_{13}$, ${U_\nu}_{23}$, ${U_\nu}_{33}$ are all real,
 and ${U_l}_{11}$, ${U_l}_{21}$, ${U_l}_{31}$ are function of $x_{l}=\frac{m_{e}}{m_{\nu}}$, $y_{l}=\frac{m_{\mu}}{m_{\tau}}$. Since, the charged lepton masses are already known, the ratios $x_{l}$ and $y_{l}$ have definite values, and so the values of ${U_l}_{11}$, ${U_l}_{21}$, ${U_l}_{31}$ are also known. 
 Hence, after further simplification of equations (\ref{eqn-imv}) and (\ref{eqn-v13}), 
 we obtain 
 \begin{equation}
  J_{CP}\sim {U_l}_{11}{U_\nu}_{13}\sin\theta+{U_l}_{21}{U_\nu}_{23}\sin\phi+
  {U_l}_{31}{U_\nu}_{33}\sin\psi.
  \label{eqn-jcp2}
 \end{equation}
Leading from the definition of $x_{\nu}$ and $y_{\nu}$ given in 
Eq.(\ref{eqn-xyn}) and combining equations (\ref{eqn-Ahat}), 
(\ref{eqn-Bhat}), and (\ref{eqn-Chat}),
the parameters 
${U_\nu}_{13}$, ${U_\nu}_{23}$, ${U_\nu}_{33}$ can be written as 
functions of the two arbitrary CP violating phases $\gamma_{\nu}$ 
and $\beta_{\nu}$ (introduced in Eq.(\ref{eqn-omnu})) as 
${U_\nu}_{13}\rightarrow g_{1}\left(\gamma_{\nu},\beta_{\nu}\right)$, 
${U_\nu}_{23}\rightarrow g_{2}\left(\gamma_{\nu},\beta_{\nu}\right)$ and 
${U_\nu}_{33}\rightarrow g_{3}\left(\gamma_{\nu},\beta_{\nu}\right)$. Hence, 
\begin{equation}
 J_{CP}\sim {U_l}_{11}g_{1}\left(\gamma_{\nu},\beta_{\nu}\right)\sin\theta
 +{U_l}_{21}g_{2}\left(\gamma_{\nu},\beta_{\nu}\right)\sin\phi+{U_l}_{31}g_{3}
 \left(\gamma_{\nu},\beta_{\nu}\right)\sin\psi.
  \label{eqn-jcp3}
\end{equation}
The functions $g_{1}\left(\gamma_{\nu},\beta_{\nu}\right)$, $g_{2}\left(\gamma_{\nu},
\beta_{\nu}\right)$, $g_{3}\left(\gamma_{\nu},\beta_{\nu}\right)$ are given by,
\begin{equation}
 g_{1}\left(\gamma_{\nu},\beta_{\nu}\right)=r\left(f_{1}-f_{2}\right)\left[\frac{1}
 {2f_{1}f_{2}(f_{1}+f_{2}-2r^{2})}\right]^{\frac{1}{2}},
\end{equation}
\begin{equation}
 g_{2}\left(\gamma_{\nu},\beta_{\nu}\right)=\left[\frac{f_{2}(f^{2}_{1}-f^{2}_{2})}
 {2f_{1}f_{2}(f_{1}+f_{2}-2r^{2})}\right]^{\frac{1}{2}},
\end{equation}
\begin{equation}
 g_{2}\left(\gamma_{\nu},\beta_{\nu}\right)=\left(f_{1}+f_{2}\right)\left[\frac{\left(f_{2}-r^{2}\right)}
 {2f_{1}f_{2}\left(f_{1}+f_{2}-2r^{2}\right)}\right]^{\frac{1}{2}},
\end{equation}
where
\begin{equation}
 f_{1}\rightarrow f_{1}\left(\gamma_{\nu},\beta_{\nu}\right)=\left[1+\hat{A}^{2}_{\nu}
 +4\left(r^{2}+\hat{B}^{2}_{\nu}+\hat{C}^{2}_{\nu}\right)+2\hat{A}_{\nu}\cos\gamma_{\nu}
 +8r\hat{B}_{\nu}\cos\beta_{\nu}\right]^{\frac{1}{2}},
 \end{equation}
and 
\begin{equation}
 f_{2}\rightarrow f_{2}\left(\gamma_{\nu}\right)=\left[1+\hat{A}^{2}_{\nu}
 +2\hat{A}_{\nu}\cos\gamma_{\nu}\right]^{\frac{1}{2}}.
\end{equation}
\section{Numerical solution of Boltzmann Equations}
\label{app:BE-sol}
\begin{figure*}[h!]
 \centering
   \begin{subfigure}{0.33\linewidth}
   \includegraphics[width=\linewidth]{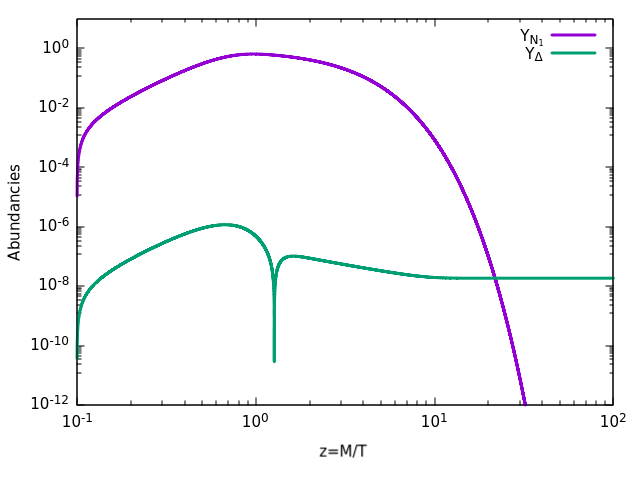}
   \caption{Set-I: $M=4\times 10^{11}$ GeV}
   \label{fig-Ia}
   \end{subfigure}
   \begin{subfigure}{0.33\linewidth}
   \includegraphics[width=\linewidth]{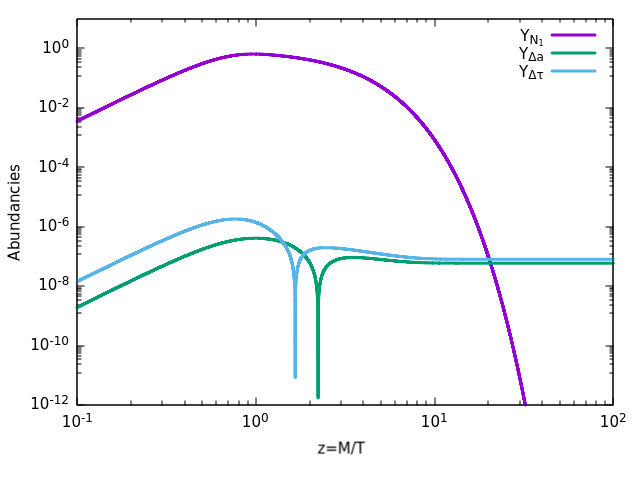}
   \caption{Set-I: $M=4\times 10^{11}$ GeV}
   \label{fig-Ib}
   \end{subfigure}
   \begin{subfigure}{0.33\linewidth}
   \includegraphics[width=\linewidth]{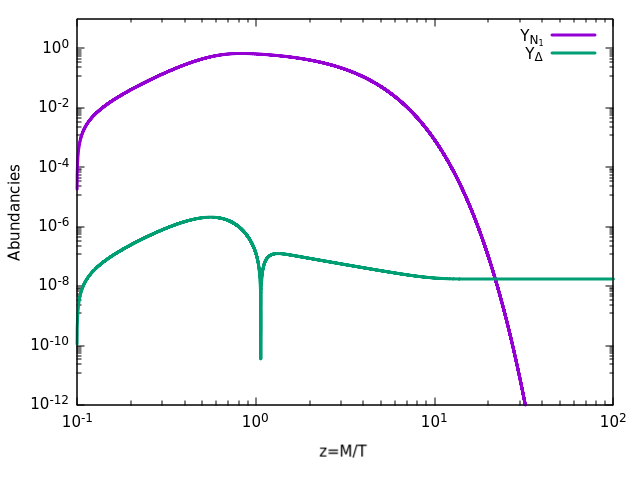}
   \caption{Set-II: $M=2\times 10^{11}$ GeV}
   \label{fig-IIa}
   \end{subfigure}
    \begin{subfigure}{0.33\linewidth}
   \includegraphics[width=\linewidth]{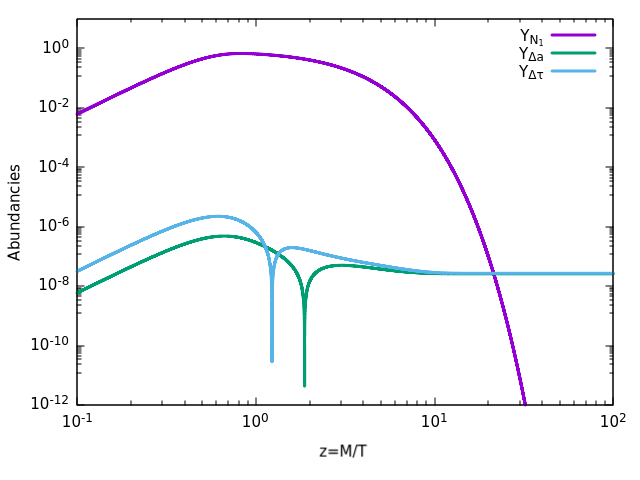}
   \caption{Set-II: $M=2\times 10^{11}$ GeV}
   \label{fig-IIb}
   \end{subfigure}
    \begin{subfigure}{0.33\linewidth}
   \includegraphics[width=\linewidth]{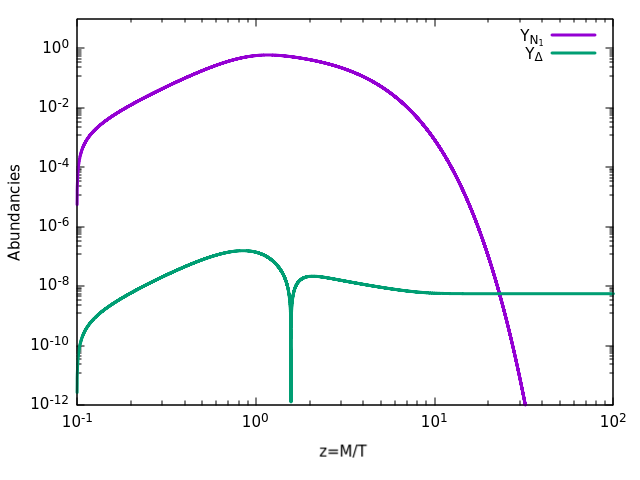}
   \caption{Set-III: $M=8\times 10^{10}$ GeV}
   \label{fig-IIIa}
   \end{subfigure}
    \begin{subfigure}{0.33\linewidth}
   \includegraphics[width=\linewidth]{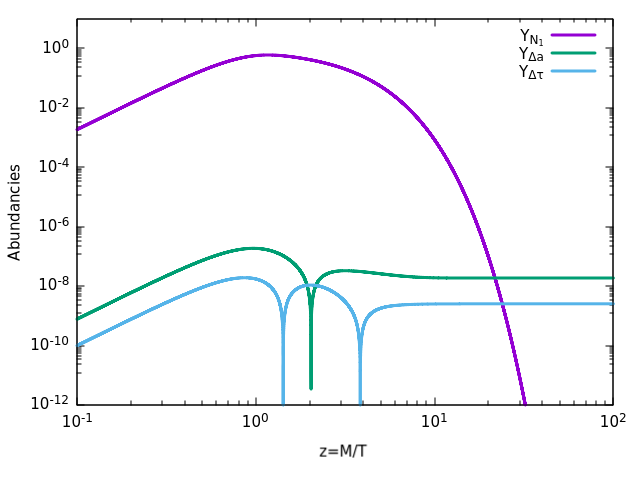}
   \caption{Set-III: $M=8\times 10^{10}$ GeV}
   \label{fig-IIIb}
   \end{subfigure}
   \begin{subfigure}{0.33\linewidth}
   \includegraphics[width=\linewidth]{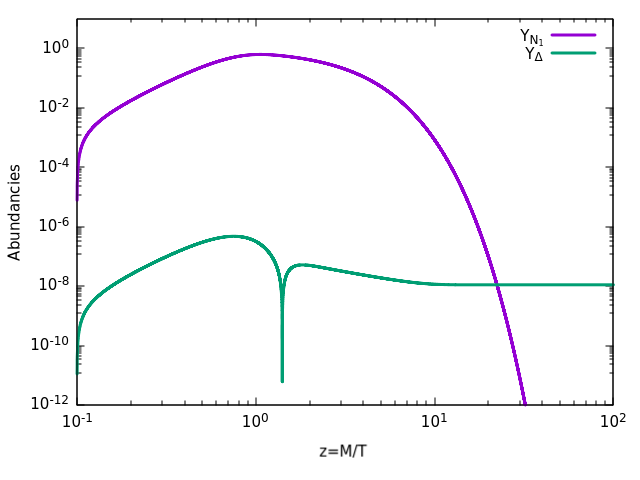}
   \caption{Set-IV: $M=4\times 10^{10}$ GeV}
   \label{fig-IVa}
   \end{subfigure}
   \begin{subfigure}{0.33\linewidth}
   \includegraphics[width=\linewidth]{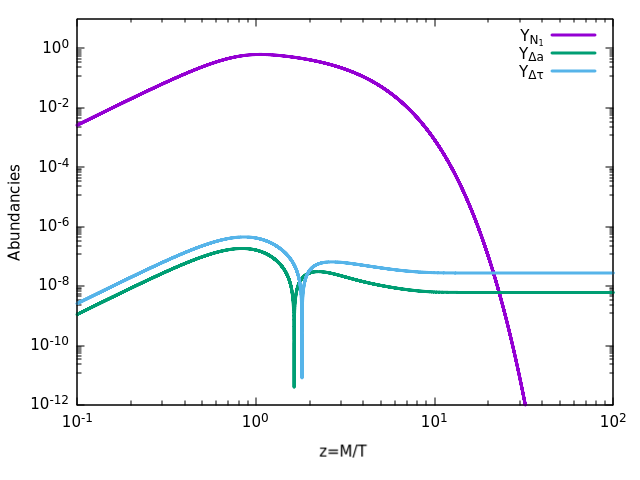}
   \caption{Set-IV: $M=4\times 10^{10}$ GeV}
   \label{fig-IVb}
   \end{subfigure}
   \caption{Evolution of 
   lepton asymmetries for
   different choices of right-handed neutrino mass $M$, in unflavoured (left column) and two-flavoured (right column) leptogenesis regimes.}
   \label{fig-fbl1}
   \end{figure*}

The solutions of the set of Boltzmann equations are depicted in the figures below (Fig.(\ref{fig-fbl1})).

\newpage
%

\end{document}